\documentclass{aa}  
\usepackage{orcidlink}
\usepackage{graphicx}
\usepackage{txfonts}
\usepackage{lipsum}
\usepackage{subcaption} 
\usepackage{multirow}

\usepackage{lscape}            
\usepackage{placeins}           
\usepackage{tablefootnote}
\usepackage{threeparttable}  
                          
\begin{document}

   \title{Environmental dependence of the properties of H$\alpha$ emitters in the extremely massive galaxy protoclusters BOSS1244 and BOSS1542 at $z = 2.24$}

   \author{Shi, Dong Dong\orcidlink{0000-0002-4314-5686}\thanks{\email{ddshi@aust.edu.cn}}\inst{\ref{aff1}}, Zheng, Xian Zhong\thanks{\email{xzzheng@sjtu.edu.cn}}\inst{\ref{aff2}}, Cai, Zheng\orcidlink{0000-0001-8467-6478}\inst{\ref{aff3}}, Fan, Xiaohui\inst{\ref{aff4}}, Wang, Xin\orcidlink{0000-0002-9373-3865}\inst{\ref{aff5}\ref{aff6}\ref{aff7}},  Dannerbauer, Helmut\inst{\ref{aff8}\ref{aff9}}, An, Fang Xia\orcidlink{0000-0001-7943-0166}\inst{\ref{aff10}}, Pan, Zhizheng\inst{\ref{aff11}}, Zhang, Yuheng\inst{\ref{aff12}\ref{aff13}},  Wen, Run\inst{\ref{aff2}}, Liu, Shuang\inst{\ref{aff14}}, Liang, Yongming\inst{\ref{aff15}\ref{aff16}}
   }
        
   \institute{Center for Fundamental Physics, School of Mechanics and Optoelectronic Physics, Anhui University of Science and Technology,
Huainan, Anhui 232001, China\label{aff1}
   \and
   State Key Laboratory of Dark Matter Physics, Tsung-Dao Lee Institute, Shanghai Jiao Tong University, Shanghai 201210, China\label{aff2}
   \and 
   Department of Astronomy, Tsinghua University, Beijing 100084, China\label{aff3}
   \and 
   Steward Observatory, University of Arizona, 933 North Cherry Avenue, Tucson, AZ 85721, USA \label{aff4}
   \and
   School of Astronomy and Space Science, University of Chinese Academy of Sciences (UCAS), Beijing 100049, China\label{aff5}
   \and
   National Astronomical Observatories, Chinese Academy of Sciences, Beijing 100101, China\label{aff6}
   \and
   Institute for Frontiers in Astronomy and Astrophysics, Beijing Normal University, Beijing 102206, China\label{aff7}
   \and 
   Instituto de Astrof\'{\i}sica de Canarias (IAC), E-38205 La Laguna, Tenerife, Spain\label{aff8}
   \and
   Universidad de La Laguna, Dpto. Astrof\'{\i}sica, E-38206 La Laguna, Tenerife, Spain\label{aff9}
   \and
   Yunnan Observatories, Chinese Academy of Sciences, Yunnan 65021, China\label{aff10}
   \and
   Purple Mountain Observatory, Chinese Academy of Sciences, 10 Yuanhua Road, Nanjing 210023, China\label{aff11}
   \and
   School of Astronomy and Space Science, Nanjing University, Nanjing 210093, China\label{aff12}
   \and
   Key Laboratory of Modern Astronomy and Astrophysics (Nanjing University), Ministry of Education, Nanjing 210093, China\label{aff13}
   \and 
   School of Physics and Electronic Engineering, Huaibei Normal University, Huaibei, Anhui 235000, China\label{aff14}
   \and
   Institute for Cosmic Ray Research, The University of Tokyo, 5-1-5 Kashiwanoha, Kashiwa, Chiba 277-8582, Japan\label{aff15} 
   \and
   National Astronomical Observatory of Japan, 2-21-1 Osawa, Mitaka, Tokyo 181-8588, Japan\label{aff16} 
   }

   \titlerunning{The properties of protocluster galaxies at $z=2.24$}
   \authorrunning{Shi et al. 2026}

  \abstract{
   Protoclusters, the progenitors of galaxy clusters, provide ideal laboratories for studying how high-density environments influence galaxy formation and evolution at early cosmic epochs. It remains unclear whether environmental effects on galaxy properties over scales of $>30$\,cMpc are already in place at $z > 2$ and how these effects depend on the evolutionary stage of large-scale structures.}
   {We investigate the physical properties of H$\alpha$ emitters (HAEs) in two extremely overdense protoclusters, BOSS1244 and BOSS1542, at $z = 2.24$, and explore the environmental dependence of galaxy properties at this redshift.}  
   {We analyze deep multi-wavelength observations of the two protoclusters. HAE candidates are selected using the narrowband $H_{2}$S1 and broadband $K_{\rm s}$ filters, with contaminants removed via $Bz_{\rm s}K_{\rm s}$ and $UBz_{\rm s}$ color selections. Given limited multi-wavelength data, stellar masses are derived from mass-to-light ratio $(M/L)$, and SFRs are estimated from H$\alpha$ luminosities corrected for [N\,II] contamination and dust extinction. Local environment is quantified using the fifth-nearest neighbor surface density $\Sigma_{\rm 5th}$.} 
   {In BOSS1244, we find a clear color--density and stellar mass--density relation. HAEs in higher-density regions are significantly redder in $z_{\rm s}-K_{\rm s}$ and more massive than those in lower-density regions. In contrast, BOSS1542 shows weak or no such correlations. Red HAEs $(z_{\rm s}-K_{\rm s}>2.5)$ in both fields are preferentially located in the densest cores, have high stellar masses $(>10^{11}\,M_{\odot})$, high SFRs, and are $4-7$ times more abundant than in general fields. The irregular morphologies observed in some red HAEs are consistent with ongoing mass assembly and may be associated with dust-obscured star formation or galaxy interactions. Both protoclusters broadly follow the star-forming main sequence, but BOSS1542 exhibits a steeper slope, indicating more active star formation in massive galaxies.}  
   {Our results demonstrate that environmental effects on galaxy properties operate over scales of $\sim39$\,cMpc at $z=2.24$, but their manifestation is strongly dependent on the evolutionary state of the large-scale structure. BOSS1244 appears more evolved, with ongoing galaxy merging and possible quenching, while BOSS1542 remains in an early, filamentary, gas-rich phase.}

   \keywords{galaxy protoclusters -- galaxy formation -- galaxy evolution -- high-redshift galaxies -- environments}

   \maketitle
   \nolinenumbers

\section{Introduction}\label{sect:intro}
The cosmic environment plays a key role in regulating the growth of galaxies. In the nearby Universe, well-known relations have been established during cluster-scale assembly, including the morphology-density \citep[e.g.,][]{Visvanathan1977,Dressler1980,Bower1992,Goto2003}, color--density \citep[e.g.,][]{Tanaka2005,Baldry2006}, and star formation rate (SFR)--density relations \citep[][]{Lewis2002,Elbaz2007}. Specifically, galaxies in rich nearby cluster environments tend to exhibit more spheroid-dominated morphologies, older stellar populations, and lower SFRs relative to galaxies in general fields, suggesting that galaxies in high-density regions form earlier than mass-matched galaxies in less dense regions \citep{Thomas2005}. The tight "red sequence" in the color-magnitude diagram (CMD), characterized by lower SFRs in cluster galaxies, persists even to $z\sim2$ \citep{Willis2020}, although the SFRs of galaxies generally increase with increasing redshift \citep{Madau2014}. Understanding how these environmental effects operate across cosmic time has become a central task in modern astrophysics. Characterizing the physical properties of member galaxies in overdense environments at high redshifts can shed light on the early evolutionary pathways of galaxies and the emergence of environmental effects.

Massive protoclusters of galaxies, the progenitors of local galaxy clusters, provide ideal laboratories to explore galaxy growth in overdense environments at early cosmic epochs. They not only accelerate evolution by fueling gas but also amplify physical mechanisms such as active galactic nuclei (AGN) feedback, shock heating, and galaxy interactions \citep[][]{Overzier2016,Alberts2022}. It remains unclear whether environmental effects on galaxy properties are already in place at $z>2$ and how these effects depend on the evolutionary stage of large-scale structures.

Studies of galaxy properties in protoclusters have yielded inconsistent results. Some protoclusters host massive quiescent galaxies with suppressed star formation \citep[e.g.,][]{Kubo2021, McConachie2022,Ito2023,Shi2024,Naufal2024,Jin2024,Ito2025,Kakimoto2026}, while others show enhanced star formation associated with an excess of dust-obscured galaxies \citep[e.g.,][]{Dannerbauer2014,Wang2016,Calvi2023,Araya-Araya2024}. Merger fractions in certain protocluster environments at $z=2-3$ are higher than those in the blank field \citep[e.g.,][]{Miley2006,Lotz2013,Hine2016,Liu2023,Giddings2026}, while other protoclusters at $z>2$ do not exhibit such an enhancement \citep[e.g.,][]{Delahaye2017, Monson2021}. Gas-phase metallicity comparisons with field galaxies show enhancements \citep{Kulas2013,Shimakawa2015}, deficits \citep{Valentino2015,Sattari2021,Chartab2021,Vito2024,Yang2026}, or no differences \citep[e.g.,][]{Kacprzak2015,Perez-Martinez2023}, suggesting that dense protocluster environments at cosmic noon can profoundly influence the interstellar medium properties of their member galaxies \citep[e.g.,][]{Li2022,Zhou2025,Wang2025,Yang2026b}. In addition, some studies suggest that the AGN fraction is greater in dense environments at $z>2$ \citep{Lehmer2009,Lehmer2013,Martini2013,Alberts2016,Krishnan2017,Polletta2021,Tozzi2022,Vito2024,Shah2025,Wu2025}. However, AGN fraction for galaxies in some protoclusters is comparable to that of field galaxies \citep[e.g., in the HS1700/SSS1558 protocluster at $z=2.30/2.53$;][]{DigbyNorth2010,Macuga2019}. These discrepancies suggest that the dynamical state and overdensity factor are essential "figures of merit" for how protoclusters regulate the growth of member galaxies \citep{Liu2023,Liu2025}. Moreover, not only the global environment but also the local environment are key to shaping the physical properties of member galaxies \citep{Ren2022}. A detailed mapping of both global and local environments in a protocluster is therefore required, yet such studies remain rare at $z>2$. 

The H$\alpha$ emission line is the preferred tracer of star formation in galaxies, as it suffers less from dust attenuation and metallicity effects than the [$\rm O\small II$] emission line or the ultraviolet (UV) continuum. Currently, spectroscopically confirmed protoclusters with H$\alpha$ emitters (HAEs) at $z=2-3$ are limited \citep{Overzier2016,Overzier2019}, and the physical properties of HAEs in these environments remain largely unclear. \cite{Koyama2013a} examined HAEs in the protocluster PKS1138 (Spiderweb) at $z=2.16$ and found that HAEs are redder and more massive in dense regions. Similar results have been reported for other candidate protoclusters traced by narrow-band (NB) selected HAEs, including 4C+10.48 at $z=2.35$, 2QZ, J0200+015 and SSA13 at $z=2.23$, and PHZG237 at $z=2.16$ \citep[][]{Hatch2011,Matsuda2011,Koyama2021,Polletta2021}, implying that HAEs in overdense regions may be more evolved systems than their underdense counterparts. However, this trend does not hold for all protocluster populations. In USS1558 at $z=2.53$, the HAEs in the densest regions are more actively star-forming than those in intergroup regions \citep{Hatch2011,Shimakawa2015,Shimakawa2018a,Shimakawa2018b,Perez-Martinez2024,Daikuhara2024}, suggesting that environmental effects depend on the specific evolutionary state of each system. Therefore, characterizing the physical properties of HAEs in protocluster environments is essential for understanding cluster formation and environmental effects.

Recently, two extremely massive protoclusters, BOSS1244 and BOSS1542, at $z=2.24\pm0.02$, have been identified through the selection of HAEs using deep narrowband (NB) H$_{2}$S1 and broadband (BB) $K_{\rm s}$ imaging observations obtained with Canada-France-Hawaii Telescope (CFHT)  \citep{Zheng2021, Shi2021}. These two systems provide valuable laboratories for investigating the environmental dependence of galaxy properties during the assembly of massive structures in the early Universe.

In this work, we carry out a comprehensive analysis of the physical properties (colors, stellar masses, SFRs) of HAEs in BOSS1244 and BOSS1542 using deep multi-wavelength imaging and spectroscopic observations. We characterize the local environment via the fifth-nearest neighbor surface density, $\Sigma_{5\mathrm{th}}$, and investigate the color--density and stellar mass--density relations, as well as star-forming main sequence (SFMS), in both protoclusters. By comparing these two systems, we explore how environmental effects on galaxy properties may depend on the assembly state of large-scale structures. We find that color--density and stellar mass--density relations are already in place at $z=2.24$, but differ dramatically between the two structures, a comparison that has not been made before. The two protoclusters, the available observations, and the data are described in Section~\ref{sec:observe}. In Section~\ref{sec:analyse}, we present the main analyses. The results and discussions are in Section~\ref{sec:results} and Section~\ref{sec:discuss}, and the conclusions are summarized in Section~\ref{sec:summary}. Throughout this paper, we adopt cosmological parameters of $\Omega_{\rm M}=0.3$, $\Omega_{\Lambda}=0.7$ and $H_0=70$ km\,s$^{-1}$\,Mpc$^{-1}$, all magnitudes are presented in the AB system. We use the Salpeter initial mass function \cite[IMF:][]{Salpeter1955} throughout unless otherwise stated. At $z=2.24$, $1'$ corresponds to $0.495$ Mpc in physical scale and $1.602$ Mpc in comoving scale, respectively.

\section{Observations, data and sample selection} \label{sec:observe} 

\subsection{BOSS1244 and BOSS1542} \label{sec:pcs}

BOSS1244 and BOSS1542 at $z=2.24\pm0.02$ were originally discovered using the MAMMOTH (MApping the Most Massive Overdensity Through Hydrogen) technique \citep{Cai2016}, which traces groups of coherently strong intergalactic Ly$\alpha$ absorption (CoSLA). They were subsequently confirmed as two extremely overdense protoclusters through the identification of HAEs selected from CFHT/WIRCam H$_{2}$S1 NB and  $K_{\rm s}$ BB imaging data \citep{Zheng2021}, together with follow-up NIR multi-object spectroscopy \citep{Shi2021}. The galaxy overdensity $(\delta_{\rm g})$ in BOSS1244/BOSS1542 over the scale of $\sim39$\,cMpc is $5.6\pm0.3 / 4.9\pm0.3$, corresponding to a total enclosed mass of $\sim10^{16} M_{\odot}$ for each structure, expected to be virialized at $z\sim0$, making them the most massive protoclusters discovered to date at $z=2-3$ \citep{Zheng2021,Shi2021}. Below, we describe these two protoclusters separately, presenting BOSS1244 first and BOSS1542 second to facilitate a direct comparison of their properties and member populations.

BOSS1244: A total of 244 HAE candidates are selected over an effective area of 417\,arcmin$^2$ using rest-frame EW$>45$ \AA\ and H$_{2}$S1$<22.5$\,mag \citep{Zheng2021}. NIR spectroscopy of 46 objects, including 41 HAEs and five quasars, obtained with MMT/MMIRS and LBT/LUCI revealed three distinct kinematic components likely in the process of coalescence \citep{Shi2021}. These components are organized into two prominent structures: a northeast (NE) component at $z=2.246$ and a southwest (SW) component at $z=2.230$, which are separated both spatially and in redshift. Their velocity dispersions are $377\pm99$\,km\,$s^{-1}$ and $405\pm202$\,km\,$s^{-1}$, respectively. Within the SW region, two additional substructures are identified in redshift space, which are likely in the process of merging to form a larger system \citep{Shi2021}. 

In addition, \cite{Zhang2022} detected 43 submillimeter sources through deep $850\,\mu$m observations with JCMT/SCUBA-2. The number density of these submillimeter sources is $2.0\pm0.3$ times higher than the average density measured in the blank fields. Importantly, these sources are spatially offset from the density peaks of the HAEs, suggesting that these extreme starbursts may be driven by physical mechanisms associated with their surrounding environments that differ from those governing the HAE population. The enhanced number density of submillimeter sources may therefore indicate that these regions have experienced intense starburst activity \citep{Zhang2022}.  

Using HST F160W imaging, \cite{Liu2023} measured the structural properties of 85 HAEs in the densest regions of BOSS1244. They found that the merger rate of HAEs is 1.8 times higher than that in coeval field environments, suggesting that the high galaxy density and dynamically young state of the protocluster may promote galaxy mergers \citep{Liu2023, Liu2025,Golden-Marx2026}. Interestingly, a pair of massive quiescent galaxies (QGs) with clear Balmer absorption features was detected at the center of the BOSS1244 SW region using \textit{Hubble Space Telescope} (HST) grism data \citep{Shi2024}. The two QGs are likely in the process of merging and are the brightest galaxies among the 40 spectroscopically confirmed member galaxies at $2.223<z<2.255$. They are expected to eventually merge and form a massive brightest cluster galaxy (BCG) with a size and stellar mass comparable to those of the most massive BCGs in the local Universe \citep{Shi2024}. Moreover, a prominent positive gradient in galaxy star formation intensity was found from the center toward the outskirts of the BOSS1244 SW region, indicating that galaxies may be undergoing gradual star formation quenching from the center outward \citep{Shi2024,Golden-Marx2026}. 

BOSS1542: This protocluster contains 223 HAE candidates over an effective area of 399\,arcmin$^2$, selected using the same criteria as those adopted for BOSS1244 \citep{Zheng2021}. NIR spectroscopy of 36 objects, including 33 HAEs and three quasars, obtained with MMT/MMIRS and LBT/LUCI, confirmed that BOSS1542 is dominated by an extended filamentary structure with a low velocity dispersion of $247\pm32$ km s$^{-1}$ at $z=2.241$ \citep{Shi2021}. This distinct kinematic configuration suggests that BOSS1244 and BOSS1542 are in different assembly stages in the formation of massive galaxy clusters \citep{Shi2021}.  

Deep $850\,\mu$m observations with JCMT/SCUBA-2 identified 54 submillimeter sources in BOSS1542 \citep{Zhang2022}. Their number density is $2.1\pm0.2$ times higher than the average density measured in the blank fields. In contrast to the HAE distribution, these sources are preferentially located away from the ridge peak of the giant filamentary structure traced by the HAEs, suggesting that violent starburst activity may be suppressed in the more dynamically evolved, collapsed regions of the growing dense structure \citep{Zhang2022}.  Furthermore, \cite{Liu2023} used the HST F160W imaging to measure the structural properties of  87 HAEs in the densest regions of  BOSS1542. The merger rate of HAEs is 2.8 times higher than that in coeval field environments and is also higher than the corresponding merger rate measured in BOSS1244, suggesting a stronger role of galaxy interactions in the extended filamentary structure of BOSS1542 \citep{Liu2023, Liu2025,Golden-Marx2026}. 

Taken together, BOSS1244 and BOSS1542 represent two particularly interesting protocluster systems for investigating the environmental effects on galaxy properties at cosmic noon.  Despite having similar redshifts and comparably high galaxy overdensities, they exhibit markedly different large-scale morphologies and dynamical states. Their contrasting properties provide an opportunity to investigate how the evolution of galaxies is linked to the assembly state of large-scale structures.  

\begin{figure*}[!ht]
\setlength{\abovecaptionskip}{0pt}
\begin{center}
\includegraphics[trim=0mm 0mm 0mm 0mm,height=0.46\textwidth]{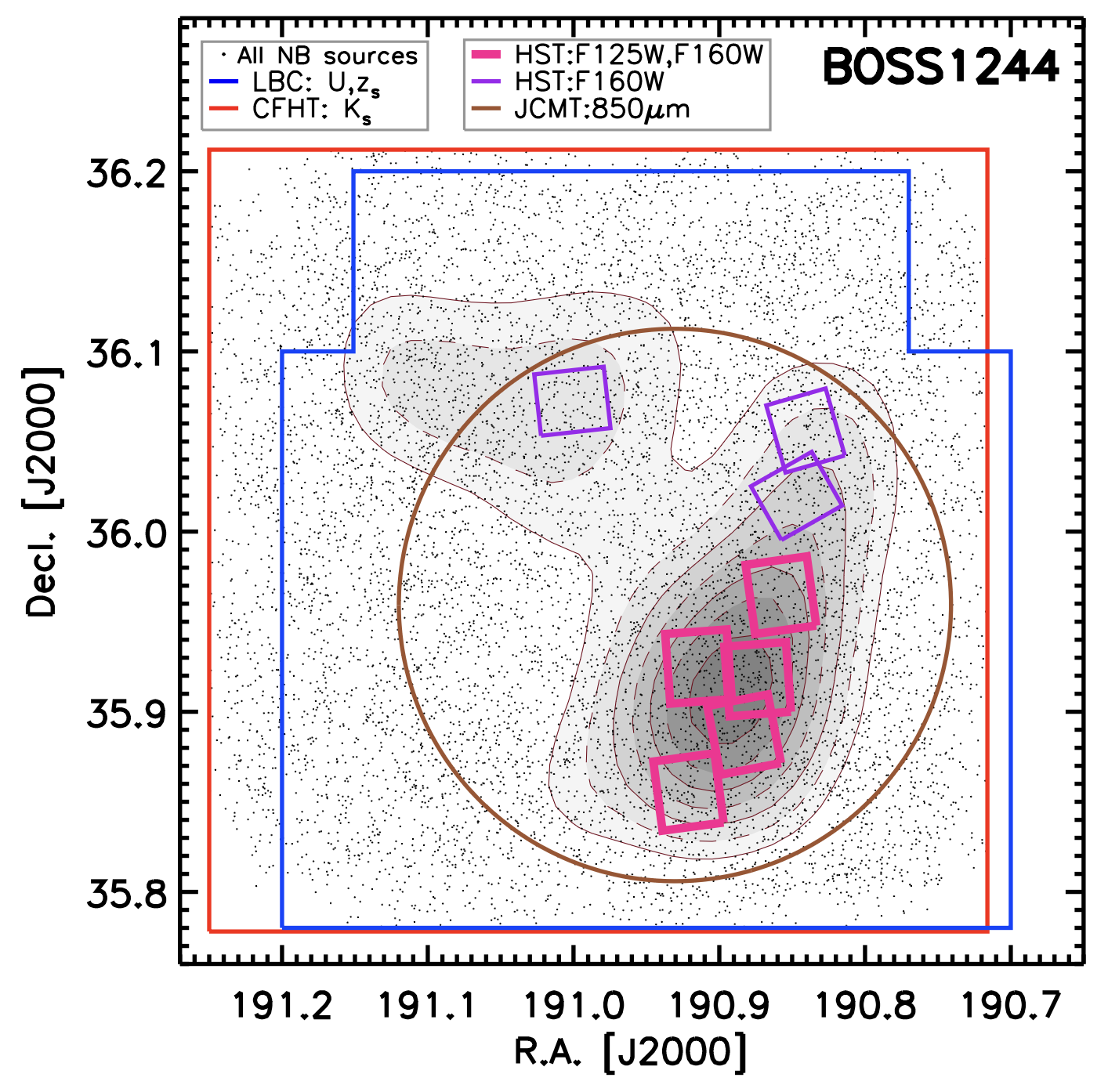}
\includegraphics[trim=0mm 0mm 0mm 0mm,height=0.46\textwidth]{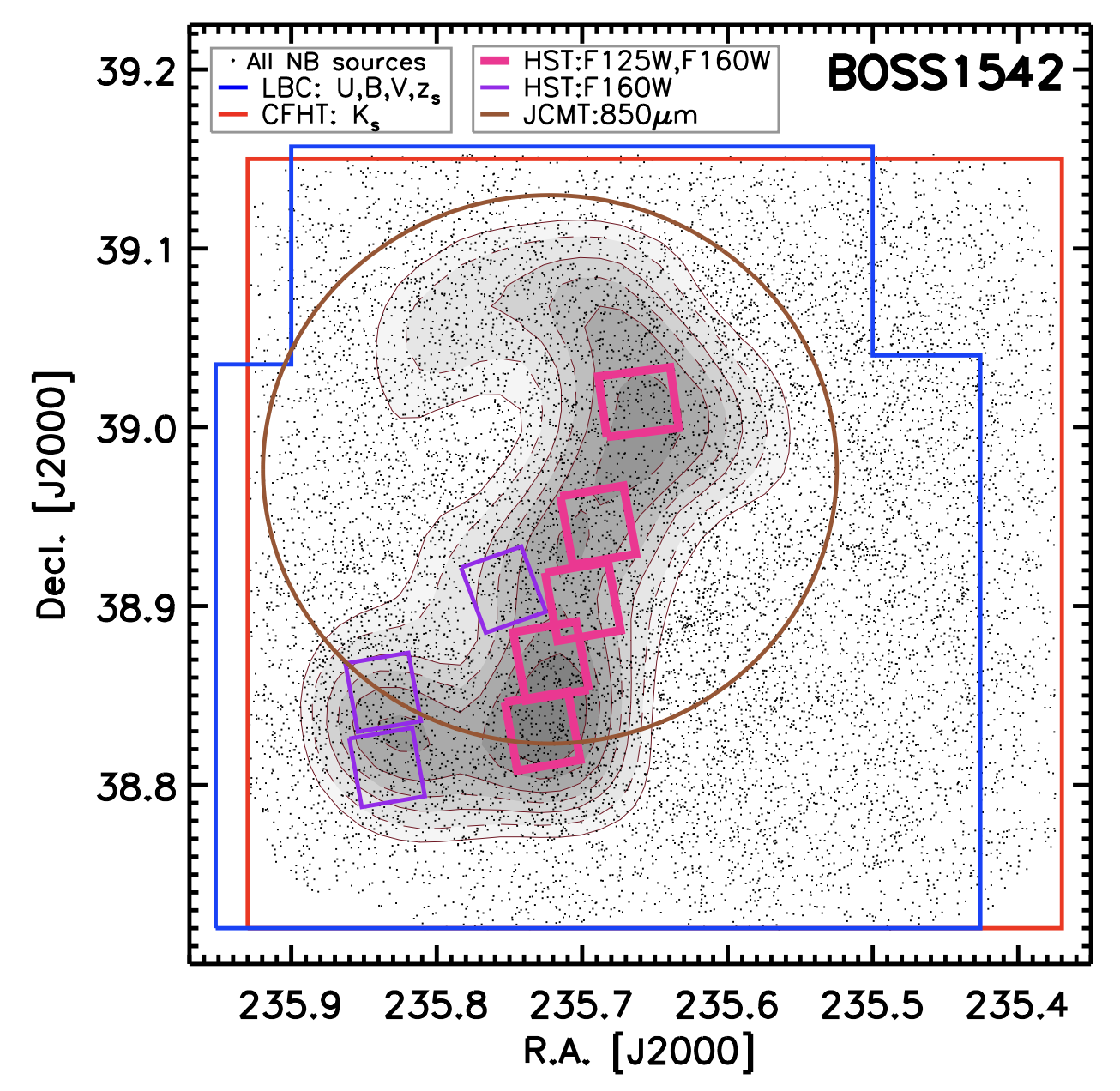}
\caption{Layout of imaging observations in BOSS1244 (left) and BOSS1542 (right). The contours in each panel show the density map of HAEs following \cite{Shi2021}. The background points denote the H$_2$S1-detected sources with CFHT. The black line outlines the sky coverage of the deep LBT/LBC imaging observations, and the red box indicates the sky coverage of the CFHT/WIRCam observations. Brown circle is the sky coverage of the JCMT 850\,$\mu$m observations. The eight small boxes (pink and purple) show the footprints of the HST/WFC3 F160W imaging observations. The five pink boxes, located at the density peaks of both fields, mark the coverage of the HST grism G141 observations, which include pre-imaging from HST/WFC3 F125W. } 
\label{fig:fig0}
\end{center}
\end{figure*}

\begin{table*}
\caption{A summary of our deep imaging data in the BOSS1244 and BOSS1542 fields.}
\label{tab:tab1}
\small
\centering
\setlength{\tabcolsep}{4pt}
\begin{tabular}{c|ccccccc}
\hline
\hline 
Field & Filter & Instrument & FoV & ObsDate & Exp. time & Limit mag. & PSF FWHM \\
 & & & [arcmin$^{2}$] & & [hrs] & [$2\arcsec,5\sigma$] & [$\arcsec$] \\
\hline
\multirow{7}{*}{BOSS1244} 
 & SDT-$U_{\rm spec}$($U$) & LBC-Blue   & $\sim$26$\times$26     & 2018/05/15                & 4.65   & 26.67 & 0.96 \\
 & $z_{\rm Sloan}$ ($z_{\rm s}$) & LBC-Red    & $\sim$26$\times$26     & 2018/05/15                & 3.92   & 25.12 & 0.63 \\
 & $H_{2}$S1 & CFHT/WIRCam & 26$\times$26     & 2016/03/20                & 7.50   & 22.58 & 0.86 \\
 & $K_{\rm s}$   & CFHT/WIRCam & 26$\times$26     & 2016/04/24                & 7.50   & 23.29 & 0.85 \\
 & F125W     & HST/WFC3    & 2.3$\times$2.1$\times$5   & 2021/01/17--2021/02/21   & $0.51\times5$ & 24.95 & 0.21 \\
 & F160W     & HST/WFC3    & 2.3$\times$2.1$\times$8   & 2018/02/05--2019/02/28   & $0.73\times8$ & 24.97 & 0.24 \\
 & 850\,$\mu$m & JCMT/SCUBA2 & 263.5                    & 2018/02--2019/04          & 18.67  & 1.07 mJy beam$^{-1}$ & 14.9 \\
\hline
\multirow{9}{*}{BOSS1542} 
 & SDT-$U_{\rm spec}$($U$)      & LBC-Blue   & $\sim$26$\times$26     & 2018/05/15,16             & 7.82   & 27.02 & 0.91 \\
 & $B_{\rm Bessel}$($B$)   & LBC-Blue   & $\sim$26$\times$26     & 2018/05/15,16             & 7.08   & 26.92 & 1.12 \\
 & $V_{\rm Bessel}$($V$)   & LBC-Red    & $\sim$26$\times$26     & 2018/05/15,16             & 2.22   & 26.30 & 0.75 \\
 & $z_{\rm Sloan}$ ($z_{\rm s}$)    & LBC-Red    & $\sim$26$\times$26     & 2018/05/15,16             & 8.97   & 25.39 & 0.61 \\
 & $H_{2}$S1  & CFHT/WIRCam & 26$\times$26     & 2016/05/21                & 7.65   & 22.67 & 0.82 \\
 & $K_{\rm s}$    & CFHT/WIRCam & 26$\times$26     & 2016/04/23                & 7.75   & 23.23 & 0.85 \\
 & F125W      & HST/WFC3    & 2.3$\times$2.1$\times$5   & 2021/02/26--2022/04/16   & $0.51\times5$ & 24.86 & 0.22 \\
 & F160W      & HST/WFC3    & 2.3$\times$2.1$\times$8   & 2019/03/27--2019/06/26   & $0.73\times8$ & 24.87 & 0.25 \\
 & 850\,$\mu$m & JCMT/SCUBA2 & 263.7                    & 2018/02--2019/04          & 21.33  & 0.98 mJy beam$^{-1}$ & 14.9 \\
\hline
\hline
\end{tabular}
\end{table*}

\subsection{Optical, NIR and submillimeter imaging data} \label{sec:optical}

Deep optical, NIR and submillimeter imaging data were taken in the BOSS1244 and BOSS1542 fields from 2016 to 2022, including SDT-$U_{\rm spec}$, $B_{\rm Bessel}$, $V_{\rm Bessel}$ and $z_{\rm Sloan}$ (these filters are abbreviated as $UBVz_{\rm s}$ in this work) from Large Binocular Telescope (LBT) observations, NIR NB H$_{2}$S1 and $K_{\rm s}$ from CFHT, F125W and F160W from HST/WFC3, and 850\,$\mu$m from JCMT/SCUBA-2. Figure~\ref{fig:fig0} represents the sky coverage of these deep imaging observations. Table~\ref{tab:tab1} summarizes our detailed imaging observations. In the following, we briefly introduce the optical, NIR, and submillimeter imaging observations and data reduction methods.

LBC: Deep optical imaging observations of the BOSS1244 and BOSS1542 fields were taken with Large Binocular Cameras (LBC) in 2018A (PI: X. Fan).  The LBCs are two wide-field imagers mounted on the prime focus ($f/1.14$) swing arms of the LBT, providing a field of view (FoV) of $\sim25\arcmin \times 23\arcmin$ and a pixel scale of $\sim0\farcs225$ \citep{Giallongo2008,Speziali2008}.  The filters $UBVz_{\rm s}$ were used for our imaging observations under the seeing conditions of $<1\farcs2$. The individual exposure times were 380\,s, 380\,s, 170\,s, and 110\,s for the $U$, $B$, $V$, and $z_{\rm s}$-band, respectively. We used the standard LBC data reduction procedures to reduce our data\footnote{\url{https://scienceops.lbto.org/data-access/data-reduction-pipelines/}}, including bias subtraction, flat-fielding, and removal of bad/hot pixels.  Source detection and photometry were performed using the SExtractor \citep{Bertin1996}. SCAMP \citep{Bertin2006} was used to perform the astrometric calibration using bright stars from SDSS, giving an accuracy of  rms $\le 0\farcs06$ in these filters. We selected the good-quality science images and stacked those of the same filter together using SWARP \citep{Bertin2002}. The bright stars of $12<u<19\,$mag with photometry from SDSS are adopted as the reference stars for photometric calibration. The limiting magnitudes ($5\sigma$, $2\arcsec$ aperture) of mosaics for these bands in the BOSS1244 and BOSS1542 fields are listed in Table~\ref{tab:tab1}.  

CFHT: Deep NIR imaging of the BOSS1244 and BOSS1542 fields in the NB H$_{2}$S1 and BB $K_{\rm s}$ filters were performed with the wide-field infrared camera (WIRCam) on CFHT during the 2016A semester (PI: FX An). WIRCam provides the FoV of $\sim20\arcmin \times 20\arcmin$ with a sampling of $0\farcs3$\,pixel$^{-1}$.  The total exposure time is 7.50\,hours for each of H$_{2}$S1 and $K_{\rm s}$ in BOSS1244, and 7.65\,hours for H$_{2}$S1 and 7.75\,hours for $K_{\rm s}$ in BOSS1542 under the seeing of $<1\farcs0$. The details of the observations and data reduction were presented in \cite{Zheng2021}. The $5\sigma$ depths within $2\arcsec$ aperture are H$_{2}$S1$= 22.58$\,mag and $K_{\rm s} = 23.29$\,mag for BOSS1244, and H$_{2}$S1$ = 22.67$\,mag and $K_{\rm s} = 23.23$\,mag for BOSS1542 (see Table~\ref{tab:tab1}). The HAE galaxy sample was selected using this instrument.

HST: Deep NIR imaging covering the density peak regions of the BOSS1244 and BOSS1542 fields was observed using the HST wide-field camera 3 (WFC3) with one-orbit in the F160W band (GO proposal 15266, PI: Z. Cai). Each pointing in HST WFC3 provides the FoV of $2\farcs1\times2\farcs3$\,arcmin$^{2}$. Eight different pointings had been performed in each field. The exposure time for each pointing is 2,614\,s. We reduced the data following the standard pipeline using ASTRODRIZZLE, and created the final NIR image mosaics with a pixel scale of 0\farcs06\,pixel$^{-1}$. The $5\sigma$ depth within $2\arcsec$ aperture in F160W are 24.97\,mag and 24.87\,mag for the BOSS1244 and BOSS1542 fields, respectively.  Additionally, five out of eight pointings in each field have HST/WFC3 F125W observations \citep{Wang2022}. The total exposure time of each pointing in F125W is 1,817\,s and the image depths ($5\sigma$, point sources) in F125W, estimated from photometry within a $2\arcsec$ aperture on random positions in the blank background regions, are 24.95\,mag and 24.86\,mag for the BOSS1244 and BOSS1542 fields, respectively. Detailed information on HST observations is provided in Table~\ref{tab:tab1} and the footprints of eight pointings in each field are shown in Figure~\ref{fig:fig0}.

JCMT: Deep 850\,$\mu$m observations of BOSS1244 and BOSS1542 were carried out with the SCUBA-2 instrument on JCMT. The PONG900 observing mode (a circle with a 15\,arcmin diameter) was adopted to scan the fields. The effective integration times are 18.67\,hours and 21.33\,hours for BOSS1244 and BOSS1542, respectively. Detailed data reduction was presented in \cite{Zhang2022}. The central noise levels reach to 1.07  and 0.98\,mJy\,beam$^{-1}$ for BOSS1244 and BOSS1542, respectively. 

\subsection{Spectroscopically confirmed HAEs} \label{sec:spec} 

We use NIR spectroscopy with MMT/MMIRS and LBT/LUCI to confirm a subsample of HAEs in BOSS1244 and BOSS1542 \citep{Shi2021}. In BOSS1244, these observations identified 18 HAEs and two quasars in the  NE region at $z=2.229-2.260$, and 23 HAEs and three quasars in the SW region at $z=2.213-2.255$. HST grism observations cover the densest regions of both BOSS1244‑SW and BOSS1542, as indicated by the pink boxes in Figure~\ref{fig:fig0}. For BOSS1244, the grism data yielded 39 secure redshifts over $z=2.15-2.35$ \citep{Wang2022}. Cross-matching these grism sources with the HAE catalog identified 14 additional HAEs in BOSS1244-SW at $z=2.213-2.246$. Combining the NIR spectroscopic and grism confirmations, we obtain a total of 18 confirmed HAEs in BOSS1244-NE and 37 HAEs in BOSS1244-SW.

In BOSS1542, the NIR spectroscopic observations confirmed 33 HAEs and three quasars at $z=2.206-2.269$ \citep{Shi2021}. The HST grism observations yielded 20 secure redshifts over $z=2.15-2.35$ \citep{Yang2026}.  Cross-matching these grism sources with the HAE catalog identified four additional HAEs in the BOSS1542-Filament region at $z=2.226-2.244$. The combined spectroscopic and grism sample therefore contains 37 confirmed HAEs in the BOSS1542-Filament region.

\begin{figure*}[!ht]
\setlength{\abovecaptionskip}{0pt}
\begin{center}
\includegraphics[trim=0mm 0mm 0mm 5mm,height=0.46\textwidth]{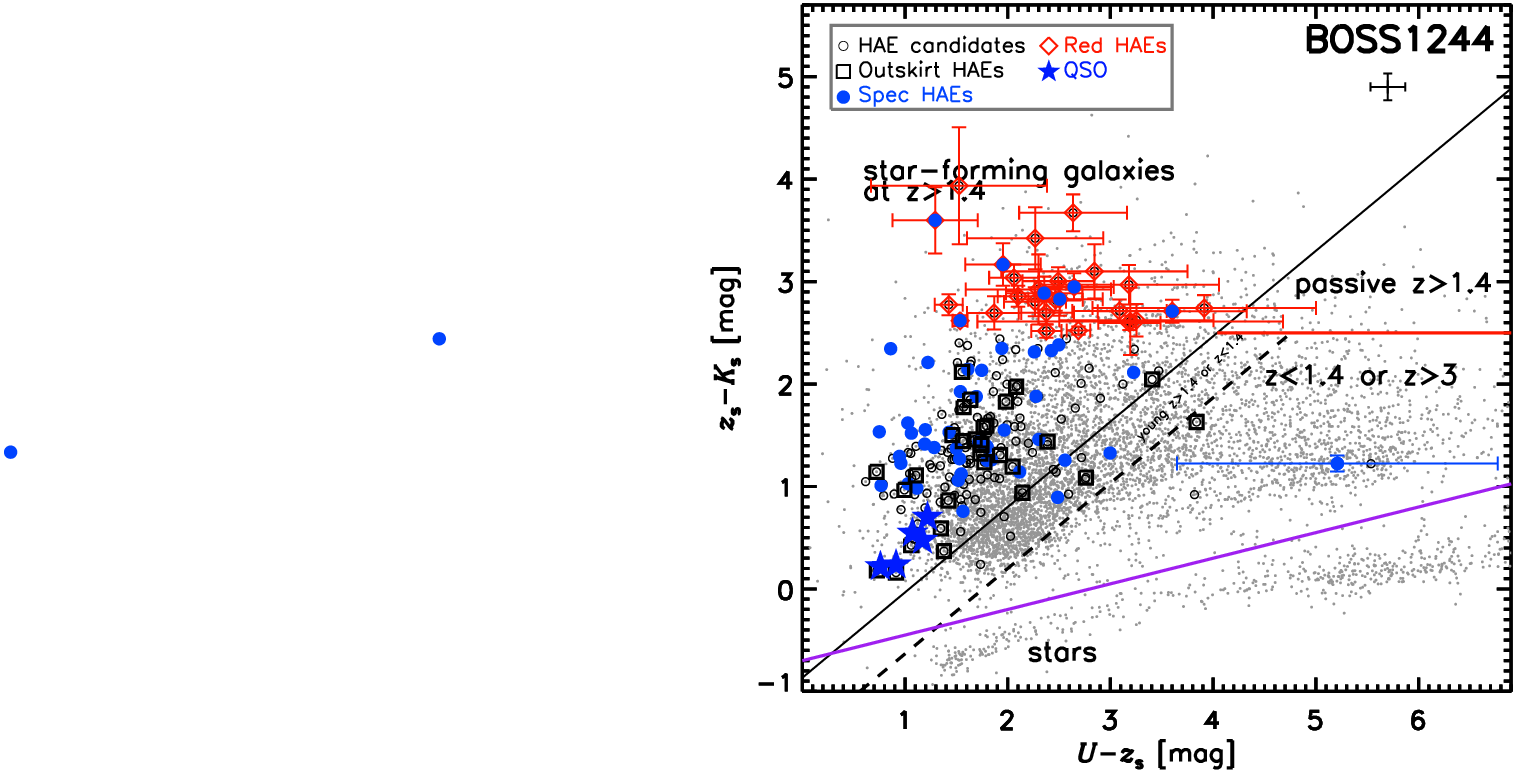}
\includegraphics[trim=0mm 0mm 0mm 5mm,height=0.46\textwidth]{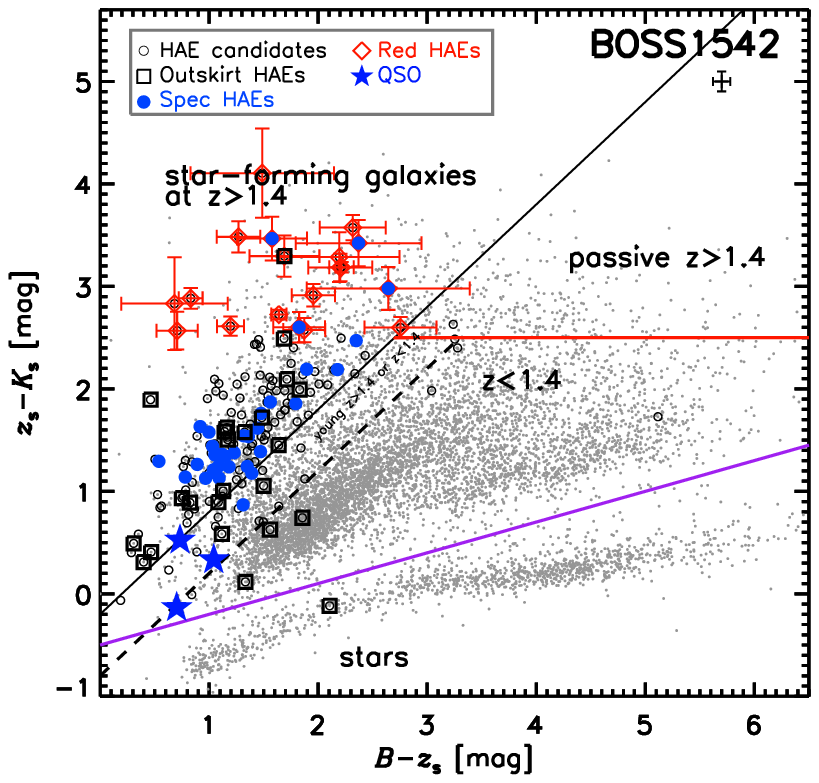}
\caption{Color selections for true HAEs at $z=2.24$ based on the $Bz_{\rm s}K_{\rm s}$ (left) and $Uz_{\rm s}K_{\rm s}$ (right) diagrams, used to remove low-redshift contaminants. Gray points represent NB-selected objects detected in $U(B)$, $z_{\rm s}$ and $K_{\rm s}$. Black solid, purple and horizontal red lines are the original $Bz_{\rm s}K_{\rm s}$ criteria from \cite{Daddi2004}. Dashed line is the modified $Bz_{\rm s}K_{\rm s}$ selection for young galaxies \citep{Lee2012}. Gray circles are the NB-selected HAEs, black squares are the HAEs in outskirt (low-density) regions of BOSS1244 and BOSS1542, and blue filled circles are the spectroscopically confirmed HAEs \citep{Shi2021}. Blue stars are the SDSS/BOSS quasars at $z=2.24\pm0.02$. Red diamonds are the red HAEs. The cross in the upper right in each panel represents the average color uncertainty of HAE galaxies. The outskirt HAEs located in underdense regions are selected as those below the level of $5.2 \times 0.071$\,per\,arcmin$^{2}$ described in \cite{Zheng2021}.} 
\label{fig:fig1}
\end{center}
\end{figure*}

\section{Analysis} \label{sec:analyse}

\subsection{Sample selection for HAEs} \label{sec:reliability}

The NB excess is typically used for HAE selection at $z\sim2.2$. The NB excess sources may contain a fraction of foreground or background contamination. Although the NB technique guarantees an excess of emission at $\lambda=2.13\,\mu$m in the selected NB emitters and is widely used to select emission line objects, this excess could also arise from  [O\,{\small III}] emitters at $z\simeq3.25$, [O\,{\small II}] emitters at $z\simeq4.71$, Pa$\alpha$/Pa$\beta$ emitters at $z\simeq0.14$/0.66, or [S\,{\small III}] emitters at $z\simeq1.23$/1.35 \citep[e.g.,][]{Geach2008, Sobral2012, An2013,An2014}. Among these (244 objects in BOSS1244 and 223 in BOSS1542), two and five HAEs in the two fields, respectively, were not covered by the LBC imaging and thus are excluded from the color-selection analysis. We select HAE candidates from the remaining samples and perform forced photometry after correcting the point spread function (PSF) matching. 

To confirm that our HAE candidates are primarily star-forming galaxies at $z > 1.4$, we apply the following criteria:  
\setlength{\abovedisplayskip}{3pt}
\setlength{\belowdisplayskip}{3pt}
\begin{equation}
\centering 
Bz_{\rm s}K_{\rm s}=(z_{\rm s}-K_{\rm s})-(B-z_{\rm s})> -0.2,
\end{equation}
which is the original criterion from \cite{Daddi2004}. We find that most of our sources satisfy this criterion, as shown in Figure~\ref{fig:fig1}.
To further account for HAE candidates dominated by young stellar populations (e.g., $\la 10{^8}\,$ yr), we extend this boundary to $-0.8$ \citep{Lee2012}:
\begin{equation}
\centering 
Bz_{\rm s}K_{\rm s}> -0.8,
\end{equation}
as shown by the black dashed line in Figure~\ref{fig:fig1}. To exclude $z > 3$ interlopers, we further apply the $UBz_{\rm s}$ criterion proposed by \cite{Hatch2011}:
\begin{equation}
\centering 
U-B<0.8+0.2(B-z_{\rm s}),
\end{equation}
Table~\ref{tab:tabnumber} summarizes the numbers of HAE at each stage of the sample-selection process in BOSS1244 and BOSS1542.

In BOSS1244, 223 HAE candidates are detected in the $Uz_{\rm s}K_{\rm s}$ bands. Since only deep $Uz_{\rm s}K_{\rm s}$ data are available for this field, we cannot directly apply $Bz_{\rm s}K_{\rm s}$ and $UBz_{\rm s}$ selections. Instead, we transform the $Bz_{\rm s}K_{\rm s}$ and $UBz_{\rm s}$ criteria onto the $Uz_{\rm s}K_{\rm s}$ color-color plane, which yields four boundary lines that define the selection window: 
 \setlength{\abovedisplayskip}{3pt}
\setlength{\belowdisplayskip}{3pt}
\begin{equation}
\centering 
z_{\rm s}-K_{\rm s}>0.833(U-z_{\rm s})-0.867 \quad \text{(black solid line)},
\end{equation}
\begin{equation}
\centering 
z_{\rm s}-K_{\rm s}>0.833(U-z_{\rm s})-1.467 \quad \text{(black dashed line)},
\end{equation}
\begin{equation}
\centering 
z_{\rm s}-K_{\rm s}=0.25(U-z_{\rm s})-0.7 \quad \text{(purle line)},
\end{equation}
\begin{equation}
\centering 
z_{\rm s}-K_{\rm s}=2.5 \quad \text{(red line)},
\end{equation}
as illustrated in the left panel of Figure~\ref{fig:fig1}. Applying these updated criteria, we reject 16 objects from the 223 HAE candidates, yielding a final sample of 207 HAEs that satisfy the $Uz_{\rm s}K_{\rm s}$ criterion in BOSS1244. Similarly, for BOSS1244, nearly all spectroscopically confirmed HAEs satisfy the $Uz_{\rm s}K_{\rm s}$ criterion, except for three HAEs that are not simultaneously detected in the $U$ and $z_{\rm s}$ bands and one HAE with large measurement uncertainties. This further demonstrates the consistency between our color selection and spectroscopic confirmation.

For BOSS1542, 180 HAEs are simultaneously detected in the $UBz_{\rm s}K_{\rm s}$ bands. Applying the $Bz_{\rm s}K_{\rm s}$ and $UBz_{\rm s}$ selection criteria rejects 29 objects.   
Consequently, 151 HAE candidates in BOSS1542 are considered to be genuine HAEs at $z=2.24$.  It should be noted that all spectroscopically confirmed HAEs satisfy both the $Bz_{\rm s}K_{\rm s}$ and $UBz_{\rm s}$ selection criteria, except for two HAEs that are not detected in the $U$ band, indicating that the color-selection criteria are highly consistent with the spectroscopic confirmation.

Based on the limited multi-band photometry, we derive photometric redshifts (phot-$z$) using the software tool EAZY \citep{Brammer2008} for $z>~2$ SFGs although the uncertainties could be large, particularly for the BOSS1244 field where imaging data are available only in the  $Uz_{\rm s}K_{\rm s}$ bands. Our results show that $\ga70\%$ of emitters have $z_{\rm phot}=1.6-2.4$ in both fields. This fraction is broadly consistent with the spectroscopic results reported by \cite{Shi2021}, which are approximately 80\% for BOSS1244 and 61\% for BOSS1542, providing independent support for the reliability of our photometric selection. We therefore regard the  galaxies satisfying the $Bz_{\rm s}K_{\rm s}$, $UBz_{\rm s}$ and $Uz_{\rm s}K_{\rm s}$ color-selection criteria as HAEs at $z=2.24\pm0.02$, while acknowledging that a small fraction of interlopers may still remain.

\begin{table}
\caption{Summary of the HAE sample selection in BOSS1244 and BOSS1542.}
\small
\label{tab:tabnumber}
\begin{tabular}{ccccc}
\hline
\hline
 Number of HAEs & BOSS1244 & BOSS1542 \\
\hline
NB-selected HAE sample  & 244  & 223 \\
Not covered by LBC imaging & 2 & 5 \\
Used for color selection & 242 & 218 \\
Detected in required color bands & 223 ($Uz_{\rm s}K_{\rm s}$) & 180 ($UBz_{\rm s}K_{\rm s}$) \\
Final color-selected HAE sample  &  207 & 151 \\
HAEs with Spec-$z$ at $z=2.24$ & 55 & 37 \\ 
Red HAEs with $z_{\rm s}-K_{\rm s}>2.5$ & 28 & 19 \\
\hline
\hline
\end{tabular}
\end{table}

\subsection{Stellar mass} \label{sec:mass} 

Stellar masses of galaxies can be well derived from spectral energy distribution (SED) fitting that incorporates rest-frame UV-to-NIR photometry. However, such rest-frame NIR data are not available for our HAE sample. \cite{Koyama2013a} demonstrated that for HAE objects, stellar masses estimated using a single-color and luminosity-based method show no systematic difference from those derived from SED fitting. Given the limited multi-band photometry available for BOSS1244 and BOSS1542, we therefore estimate the stellar mass using the more complete $K_{\rm s}$-band luminosity and the theoretical mass-to-light ratio $(M/L)$ of the galaxies. This approach is practically effective and has been widely adopted in protocluster studies at similar redshifts \citep[e.g.,][]{Hayashi2012, Koyama2013a, Koyama2014}. 

We estimate stellar mass using the equation from \cite{Koyama2013a}: 
\begin{equation}
\centering 
{\rm log}(M_{*}/10^{11}\,M_{ \odot })=-0.4(K_{\rm s}-22.24)+{\rm \Delta log} M,
\end{equation}
where we derive a simple scaling of the model galaxies with $M_{*}=10^{11}\,M_{ \odot }$ ( with $K_{\rm s}$=22.24\,mag) from \cite{Kodama1999}.
The last term is set as ${\rm \Delta\,log} M=0.03-1.50\times$\,exp[$-1.11\times (z_{\rm s}-K_{\rm s}$)], where $z_{\rm s}-K_{\rm s}$ provides a sensitive tracer of the typical age of the stellar population in these HAEs because it straddles the redshifted 4000\AA \,break \citep{Koyama2013a}. 
 
\subsection{Star formation rate} \label{sec:sfr}

H$\alpha$ luminosity is a good indicator of SFR  because the H$\alpha$ emission line originates from the photoionization by young O and B stars, thus probing instantaneous star formation activity on a timescale of $<20$\,Myr. It is well calibrated and widely used in studies of both local and high-$z$ galaxies. Moreover, it is less sensitive to dust extinction than [O\,{\small II}]  and UV luminosities. 

We compute SFRs of sample HAEs using H$\alpha$ luminosity. The H$\alpha$ fluxes are derived from H$_{2}$S1 and $K_{\rm s}$ photometry using the following equations: 
\begin{equation}
\centering
F_{\rm H\alpha+[N\small II]}=F_{\rm line},
\end{equation}

\begin{equation}
\centering
f_{\rm c}=\frac{f_{K_{\rm s}}-f_{\rm H_{2}S1}(\Delta \lambda_{\rm H_{2}S1}/\Delta \lambda_{K_{\rm s}})}{1-\Delta \lambda_{\rm H_{2}S1}/\Delta \lambda_{K_{\rm s}}},
\end{equation}
 
\begin{equation}
\centering
\rm EW_{\rm rest}=(1+z)^{-1}\frac{F_{\rm H\alpha+[N\small II]}}{f_{\rm c}},
\end{equation}
where $F\rm_{H\alpha+[N\rm \small II]}$ refers to line flux, $f_{\rm c}$ is continuum flux density and EW$_{\rm rest}$ is rest-frame equivalent width (EW) of H$\alpha$.    
$\Delta \lambda_{\rm H_{2}S1}=0.0295\,\mu$m and $\Delta\lambda_{K_{\rm s}}=0.327\,\mu$m are the full widths at half maximum (FWHMs) of the $H_{2}$S1 and $K_{\rm s}$ filters, and $f_{\rm H_{2}S1}$ and $f_{K_{\rm s}}$ are the flux density in the $\rm H_{2}S1$ and $K_{\rm s}$-band. We adopt an empirical correction for the $\rm [N\small II]\lambda\lambda6548,6583$\AA\,line contribution to NB fluxes using the relation between $\rm [N\small II]$/H$\alpha$ and the rest-frame EW$_{\rm H\alpha+[N\small II]}$. The equations are $f=\rm log (\rm [N\small II]/H\alpha)$, $E=\rm log (\rm EW_{\rm rest}) $, and $f= -0.924 + 4.802E - 8.892E^2 + 6.701E^3 - 2.27E^4 + 0.279E^5$, as presented in \cite{Sobral2009, Sobral2012}. The average correction $\rm [N\small II]/H\alpha$ for the $z = 2.24$ HAE sample is 0.16 and 0.15 for BOSS1244 and BOSS1542, respectively, with average rest-frame $EW_{\rm rest}$ of 145\,\AA\,and 148\,\AA, respectively.

We correct the H$\alpha$ fluxes of individual galaxies for dust extinction using the stellar-mass-dependent $A_{\rm H\alpha}-M_{*}$ relation established for local galaxies by \cite{Garn2010}. Although the intrinsic scatter in this relation introduces some uncertainty in the derived corrections, it remains one of the most reliable empirical dust estimators currently available. Importantly, this relation has been shown to remain valid out to $z\sim1.5-2$ \citep{Hayashi2011, Sawicki2012, Koyama2013a, Koyama2014}, which justifies its application to high-redshift samples. We therefore consistently apply this same extinction correction to all H$\alpha$ emitters in our study. 

SFRs are calculated using the relation from \cite{Kennicutt1998} assuming a \cite{Salpeter1955} IMF with the following formula:
\begin{equation}
\centering
\rm SFR\,(\rm M_{\odot }\,yr^{-1})=7.9\times10^{-42}L_{\rm H\alpha} (\rm erg\,s^{-1}) .
\end{equation}

\begin{figure*}[!ht]
\setlength{\abovecaptionskip}{0pt}
\begin{center}
\includegraphics[trim=0mm 0mm 0mm 0mm,height=0.4\textwidth]{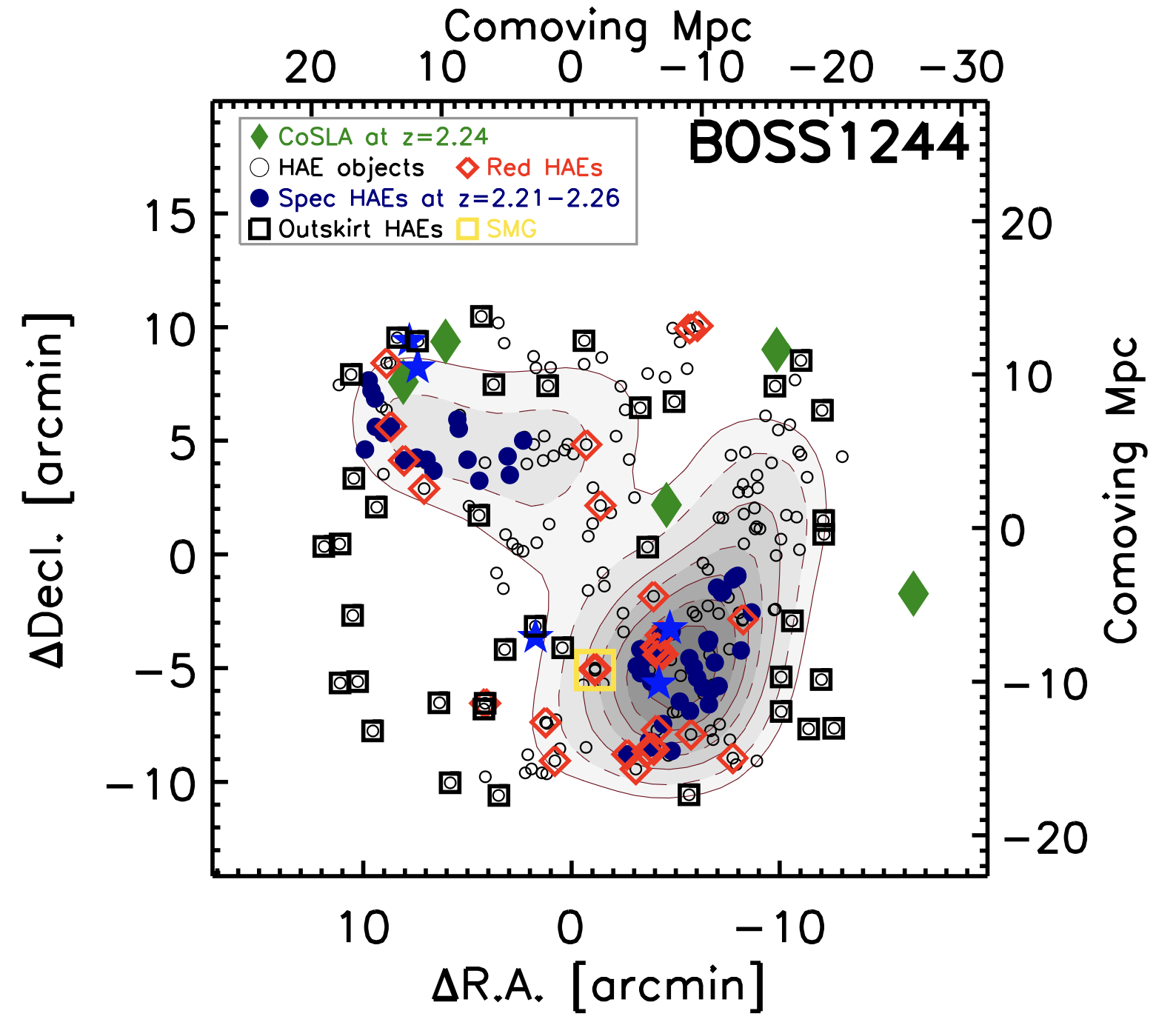}
\includegraphics[trim=0mm 0mm 0mm 0mm,height=0.4\textwidth]{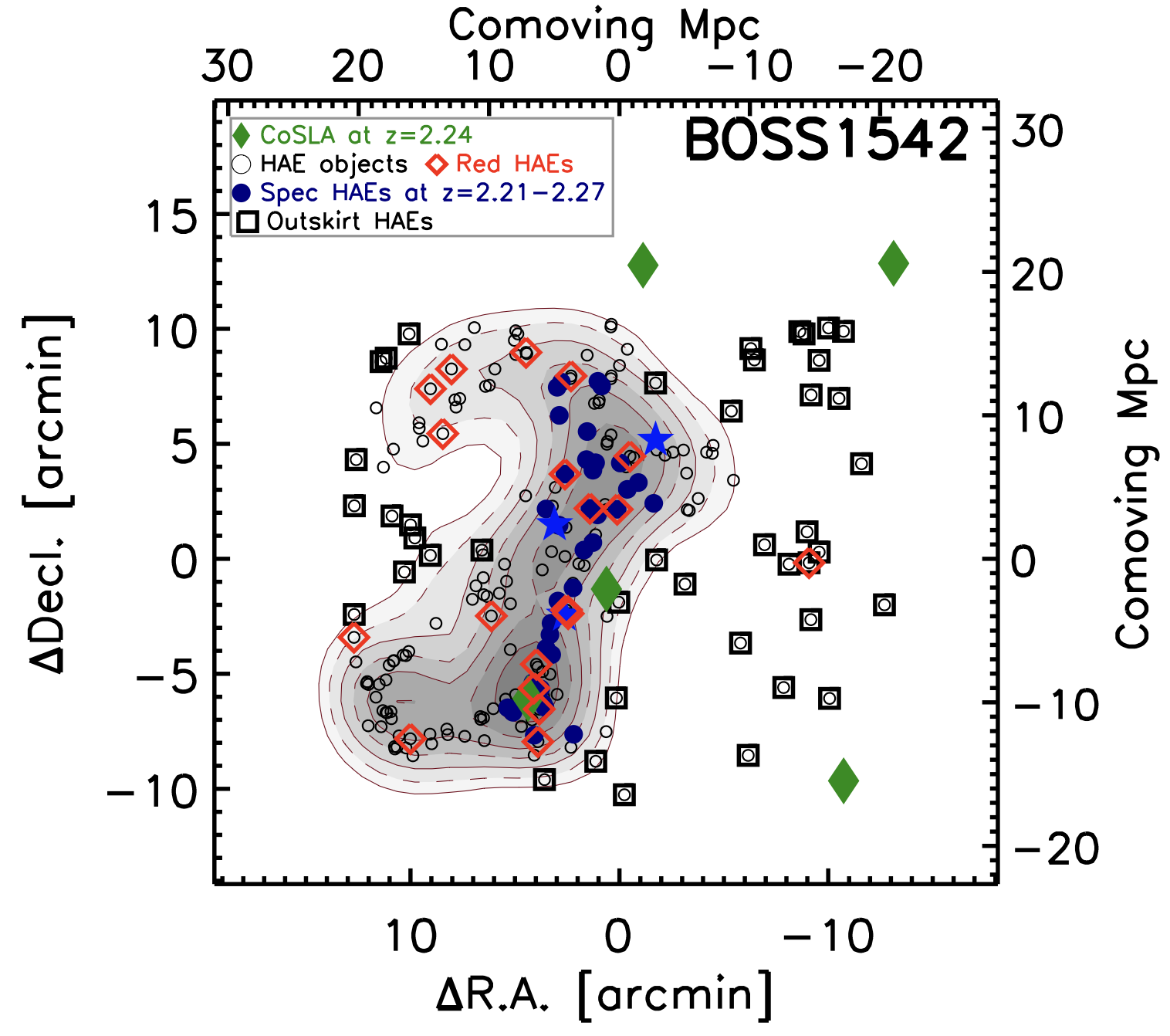}
\caption{Spatial distribution of HAEs in BOSS1244 (left) and BOSS1542 (right). The contours in each panel show the density map of HAEs following \cite{Shi2021}. Blue stars and green diamonds denote SDSS/BOSS quasars and groups of Ly$\alpha$ absorption systems at $z=2.24$, respectively. The yellow square in the left panel highlights the region where an SMG overlaps with two HAEs. Circles represent  the NB-selected HAE candidates, while blue filled circles refer to the spectroscopically confirmed HAEs at $z=2.21-2.27$. Black squares and red diamonds mark the outskirt HAEs and red HAEs with $z_{\rm s}-K_{\rm s}>2.5$, respectively. } 
\label{fig:fig2}
\end{center}
\end{figure*}

\subsection{Environment} \label{sec:environment}

To investigate the environmental dependence of galaxy properties at $z=2.24$, we use the average surface density parameter $\Sigma_{\rm 5th} = 5 / (\pi r_{\rm 5th}^{2})$ as a tracer of the local environment, where $r_{\rm 5th}$ is the projected distance to the fifth-nearest neighbor of each HAE. The measured median values of $\Sigma_{\rm 5th}$ in BOSS1244 and BOSS1542 are 3.483 and 4.130, respectively. For comparison, we also compute $\Sigma_{\rm 5th}$ for HAEs  at $z=2.23\pm0.02$ in the blank fields including COSMOS, UDS and ECDFS. These field HAEs are drawn from NB surveys \citep{Sobral2013, An2014}. Although the NB filters and detection limits are not exactly identical to those of the BOSS1244 and BOSS1542 observations (e.g., the COSMOS and UDS NB filters differ slightly, and the ECDFS detection limit is slightly deeper), we apply the same selection criteria to all field HAEs as those adopted for BOSS1244 and BOSS1542 (NB excess $\Sigma>3$, rest-frame $\rm EW>45\AA$ and $\rm NB<22.5$\,mag). The common $\rm NB<22.5\,mag$ cut also ensures a consistent magnitude limit for the selected HAEs, and the selection function does not explicitly depend on stellar mass or SFR. These field HAEs are used only to provide a field baseline for the density measurements and are not used to construct the protocluster samples.

We find that the average local surface densities in BOSS1244 and BOSS1542 are approximately 9 and 11 times those in the UDS and ECDFS fields, respectively, and about 5 and 6 times that in COSMOS. Notably, the $\Sigma_{\rm 5th}$ in COSMOS is higher than in UDS and ECDFS, which can be attributed to a known high-density structure in COSMOS reported in previous studies \citep{Geach2012, Darvish2020}. The average $\Sigma_{\rm 5th}$ values for BOSS1244, BOSS1542 and three blank fields are summarized in Table~\ref{tab:tab2}. We conclude that BOSS1244 and BOSS1542 are indeed extremely overdense on average, consistent with the number density maps presented in our earlier works \citep{Zheng2021,Shi2021}. Finally, we note that our results are not sensitive to the specific choice of environmental density tracer, as using $\Sigma_{\rm 3th}$ or $\Sigma_{\rm 10th}$ yields consistent results.

\begin{table}
\caption{The local surface density  ($\rm \Sigma_{\rm 5th}$) of HAEs in our fields and other coeval blank fields.}
\small
\label{tab:tab2}
\begin{tabular}{ccccc}
\hline
\hline
Field & Redshift & $\#$ of HAEs & Area & $\rm \Sigma_{\rm 5th,median}$ \\
 & & & [arcmin$^{2}$] & [pMpc$^{-2}$] \\
\hline
ECDFS  & 2.24 & 28  & 383  & 0.378 \\
UDS    & 2.23 & 177 & 2480 & 0.372 \\
COSMOS & 2.23 & 516 & 6338 & 0.646 \\
\hline
BOSS1244 & 2.24 & 244 & 417 & 3.483 \\
BOSS1542 & 2.24 & 223 & 399 & 4.130 \\
\hline
\hline
\end{tabular}
\end{table}

\section{Results}\label{sec:results}

BOSS1244 and BOSS1542, traced by HAEs over scales of $30-40$\,cMpc, exhibit different morphologies, suggesting that they may be at different stages of evolution \citep{Shi2021}. 
The spatial distributions of NB-selected HAEs can identify structures and substructures in and around a protocluster. Figure~\ref{fig:fig2} shows the spatial distributions of the HAE sample as well as CoSLA and quasars at the coincident redshift of $z=2.24\pm0.02$ in BOSS1244 and BOSS1542. Using NIR spectroscopic data, \cite{Shi2021} reported that BOSS1244 exhibits two prominent structures (the NE and SW components) both in redshift space and on the sky, whereas BOSS1542 displays an extended filamentary structure with a very low velocity dispersion of $247\pm32$\,km\,s$^{-1}$. Therefore, we divide BOSS1244 and BOSS1542 into BOSS1244-NE, BOSS1244-SW and BOSS1542-Filament, respectively. We use these three components to investigate how galaxy properties vary among the distinct structures. Our goal is to investigate physical properties of HAEs in these two extremely massive structures at $z=2.24$ and their environmental dependence in order to better understand the impact of the environment on the physical properties of star-forming galaxies.

\begin{figure*}[ht]
\setlength{\abovecaptionskip}{0pt}
\begin{center}
\includegraphics[trim=0mm 0mm 0mm 0mm,clip,height=0.43\textwidth]{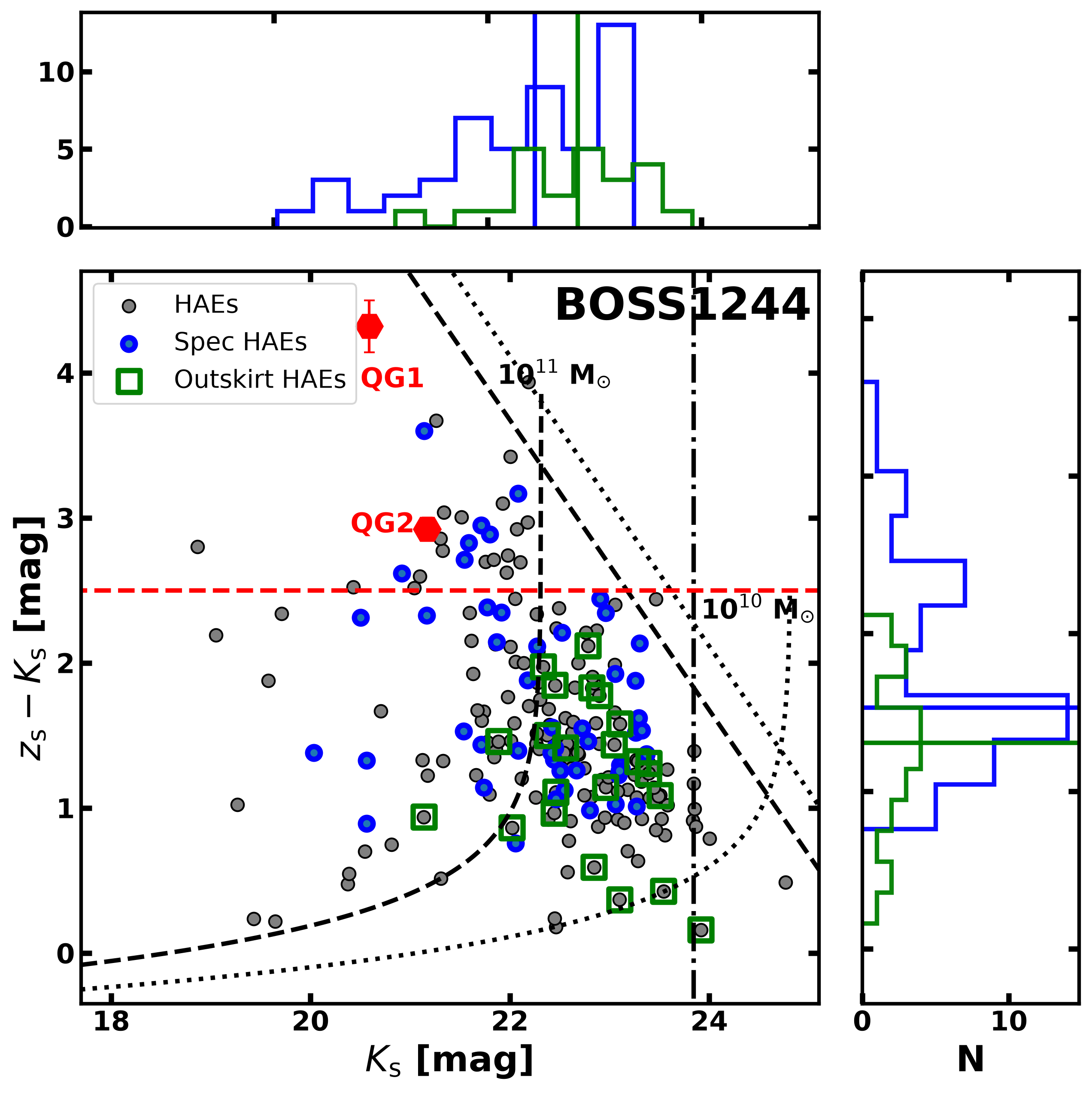}
\includegraphics[trim=0mm 0mm 0mm 0mm,clip,height=0.43\textwidth]{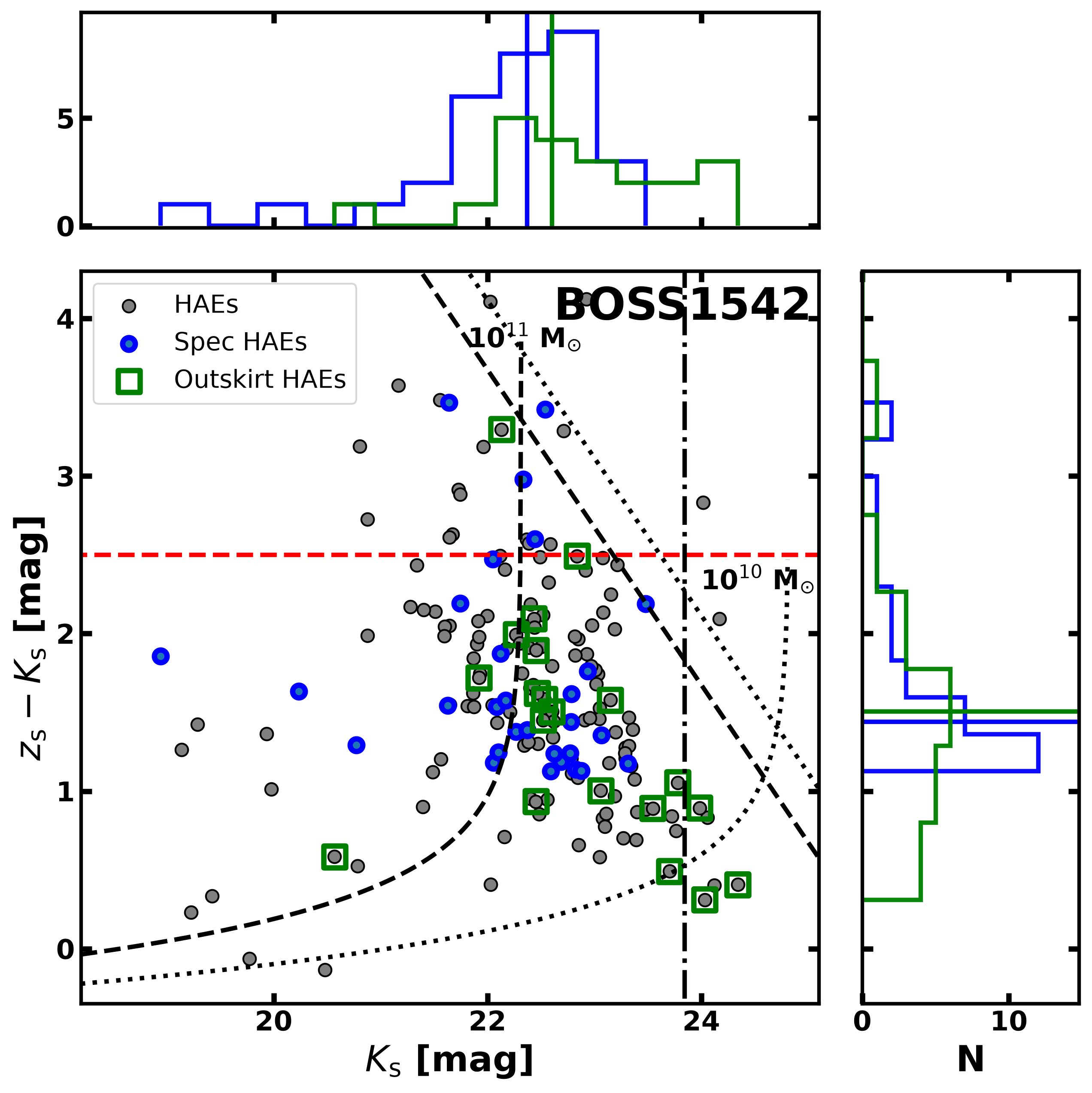}
\caption{Color–magnitude diagram for all NB-selected HAEs in BOSS1244 (left) and BOSS1542 (right). The red hexagons in the left panel mark QG1 and QG2, two massive quiescent galaxies located in the BOSS1244-SW region \citep{Shi2024}.
The gray points are the NB-selected HAEs which are detected at both $z_{\rm s}$ and $K_{\rm s}$. The slanted dashed and dotted lines show the 2$\sigma$ and 3$\sigma$ limiting magnitudes in the $z_{\rm s}$-band, respectively. The vertical dot-dashed line is the 3$\sigma$ limiting magnitude in $K_{\rm s}$. The dotted and dashed curves show the color-magnitude track of a galaxy with $M_{*}$ of $10^{10}\,M_{ \odot }$ and $10^{11}\,M_{ \odot }$, respectively. The green squares are the outskirt HAEs, and the blue circles are spectroscopically confirmed HAEs. The red dashed line indicates the color threshold $z_{\rm s}-K_{\rm s}=2.5$. The histograms show the distributions of HAE colors  $z_{\rm s}-K_{\rm s}$ and $K_{\rm s}$ magnitudes. The blue and green vertical lines indicate the median values for the spectroscopically confirmed HAEs and the HAEs in the outskirts, respectively.} 
\label{fig:fig3}
\end{center}
\end{figure*}

\subsection{Color-magnitude diagram of HAEs }\label{sec:cmd}

Figure~\ref{fig:fig3} presents the color-magnitude diagrams (CMDs) of HAEs in BOSS1244 and BOSS1542. Red circles denote spectroscopically confirmed HAEs located in high-density regions, while black squares represent outskirt HAEs inhabiting low-density regions of both systems. The spectroscopically confirmed HAEs in the high-density regions tend to be brighter in the $K_{\rm s}$-band than those in the outskirts, with a more pronounced difference in BOSS1244.  Specifically, the median $K_{\rm s}$-band magnitudes are $22.44\pm0.07$ and $22.84\pm0.09$ for the high-density and outskirt HAEs in BOSS1244, respectively, corresponding to an offset of 0.40\,mag. For BOSS1542, the corresponding median magnitudes are $22.37\pm0.07$ and $22.60\pm0.09$, respectively, giving a smaller offset of 0.23\,mag. The $K_{\rm s}$-band magnitude distributions are also shown in the top panels of Figure~\ref{fig:fig3}.

Furthermore, in the high-density region of BOSS1244, HAEs exhibit redder colors ($z_{\rm s}-K_{\rm s}=1.53\pm0.11$) than those in its underdense regions $(z_{\rm s}-K_{\rm s}=1.31\pm0.13)$. In contrast, the color of HAEs in the high-density filamentary region of BOSS1542 ($z_{\rm s}-K_{\rm s}=1.44\pm0.09$) is comparable to that of HAEs in its low-density regions ($z_{\rm s}-K_{\rm s}=1.50\pm0.14$). These differences in the environmental dependence of HAE colors may be related to the different dynamical stages of BOSS1244 and BOSS1542.

In Figure~\ref{fig:fig3}, we show color-magnitude tracks for galaxies of fixed  stellar mass $M_{*}=10^{10}\,M_{\odot }$ and $10^{11}\,M_{\odot }$ using the same model described in Section~\ref{sec:mass}. Notably, HAEs with red color in $z_{\rm s}-K_{\rm s}$ typically exhibit very high stellar masses ($M_{*} \ge 10^{11}\,M_{\odot}$). These massive, actively assembling HAEs, largely absent in low-redshift clusters, are likely the progenitors of the red sequence galaxies observed in present-day clusters. This suggests that cluster galaxies had already formed a significantly large fraction of their stars by $z\sim2.2$, even though they were still undergoing intense star formation at that epoch.

\begin{figure}[ht]
\setlength{\abovecaptionskip}{0pt}
\begin{center}
\includegraphics[trim=0mm 0mm 0mm 0mm,clip,height=0.6\textwidth]{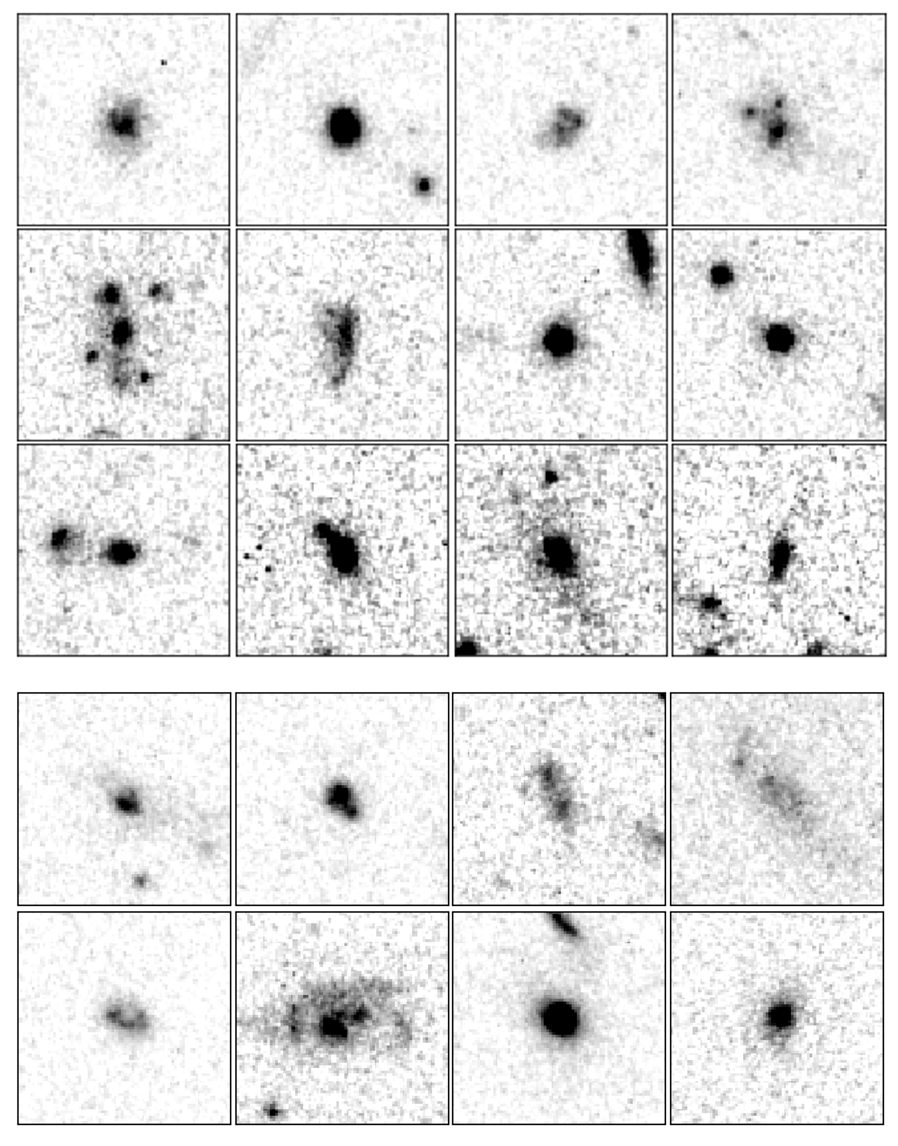}
\caption{HST/F160W cutouts of 12 red HAEs in BOSS1244 (upper) and 8 red HAEs in BOSS1542 (bottom). The size of each postage stamp image is 5\arcsec $\times$ 5\arcsec.} 
\label{fig:fig4}
\end{center}
\end{figure}

\begin{table}
\centering
\begin{threeparttable} 
\caption{Comparison of the median $K_{\rm s}$ and stellar mass ($\log M_{*}$) of red HAEs ($z_{\rm s}-K_{\rm s}>2.5$) and other HAEs ($z_{\rm s}-K_{\rm s}<2.5$) in BOSS1244 and BOSS1542.}
\label{tab:tab33}
\small
\setlength{\tabcolsep}{4pt}
\begin{tabular}{c|ccc}
\hline 
\hline
Field & Color & $K_{\rm s, med}$ & $\log(M_{*})_{\rm med}$  \\
 & [mag] & [mag] & [$M_{\sun}$]   \\
\hline
\multirow{2}{*}{BOSS1244} 
 &$z_{\rm s}-K_{\rm s}>2.5$ & $21.63\pm0.22$ & $11.20\pm0.09$ \\
 &$z_{\rm s}-K_{\rm s}<2.5$ & $22.74\pm0.11$ & $10.52\pm0.06$ \\
 \hline
\multirow{2}{*}{BOSS1542} 
&$z_{\rm s}-K_{\rm s}>2.5$ & $21.96\pm0.33$ & $11.05\pm0.13$ \\
 &$z_{\rm s}-K_{\rm s}<2.5$ & $22.69\pm0.13$ & $10.60\pm0.08$ \\
\hline
\hline
\end{tabular}
\begin{tablenotes} 
\footnotesize
\item[$\ast$] Note: $K_{\rm s, med}$ and $\log(M_{*})_{\rm med}$ represent the median values of the $K_{\rm s}$ band and stellar mass, respectively.
\end{tablenotes}
\end{threeparttable}
\end{table}

\subsection{Red HAEs}\label{sec:rhea}

In low-redshift ($z<1$) clusters, some red HAEs and dust-obscured star-forming galaxies have already been identified \citep{Geach2006, Koyama2010, Koyama2011}. These red HAEs are more likely to be dust-obscured galaxies, and are considered to be systems in transition from intense starburst activity to passively evolving quiescent galaxies. Their properties may also be influenced by environmental effects, as such red HAEs are preferentially found in intermediate-density regions of clusters at $z<0.8$, where a rapid color transition from blue to red is observed \citep{Tanaka2005, Koyama2008}. Consequently, some galaxies located at the faint end of the red sequence, built up by such transitioning systems, could also be identified as red HAEs. 
At higher redshifts, however, the nature of red HAEs may be different, particularly in the dense environments of protoclusters. For example, \cite{Koyama2013a} studied red HAEs in the $z=2.16$ PKS1138 protocluster using Spitzer-$24\,\mu$m data and found that the majority of $24\,\mu$m-detected HAEs are red. They argued that most red HAEs are luminous, dust-obscured starburst galaxies rather than quiescent systems.

We employ the color criterion $z_{\rm s}-K_{\rm s}>2.5$ to select red HAEs \citep{Hayashi2011}. In BOSS1244, we identify 28 red HAEs, seven of which have spectroscopic redshifts. We find that these galaxies are predominantly clustered in the denser regions of the protocluster, with a notable concentration in the SW area. In contrast, very few red HAEs are found in the low density outskirts. In BOSS1542, we identified 19 red HAEs, three of which have spectroscopic redshifts. Similarly, these red HAEs are mainly concentrated along the filamentary structure of BOSS1542, while their fraction in the outer regions remains very low, as indicated by the red diamonds in the right panel of Figure~\ref{fig:fig2}. 

The red HAEs in both protoclusters are systematically brighter in the $K_{\rm s}$ band and more massive than the HAEs with $z_{\rm s}-K_{\rm s}<2.5$.  We have provided a quantitative comparison of the median $K_{\rm s}$-band  and stellar masses in Table~\ref{tab:tab33}.
In BOSS1244, the median $K_{\rm s}$-band magnitudes and stellar masses are $21.63\pm0.22$ mag and log$(M_*/M_{\odot})=11.20\pm0.09$ for the red HAEs, compared with $22.74\pm0.11$ mag and $10.52\pm0.06$ for the HAEs with $z_{\rm s}-K_{\rm s}<2.5$, respectively. Similarly, in BOSS1542, the corresponding median values are $21.96\pm0.33$ mag and $11.05\pm0.13$ for the red HAEs, compared with $22.69\pm0.13$ mag and $10.60\pm0.08$ for the HAEs with $z_{\rm s}-K_{\rm s}<2.5$. These comparisons quantitatively demonstrate the systematically brighter $K_{\rm s}$-band magnitudes and higher stellar masses of the red HAEs. The physical properties of these red HAEs in BOSS1244 and BOSS1542 are listed in Table~\ref{tab:tab3} and Table~\ref{tab:tab4}. 

Using the same selection criteria (NB excess $\Sigma>3$, $\rm EW>45\,\AA$ and $\rm NB<22.5$\,mag), we identified a total of 54 HAEs at $z=2.2$ with $z_{\rm s}-K_{\rm s}>2.5$ in the 1.7\,deg$^{2}$ COSMOS field, yielding a surface density of $(8.81 \pm 1.20) \times 10^{-3}$\,arcmin$^{-2}$. HAEs in COSMOS were selected with the same criteria as those in BOSS1244 and BOSS1542, and are used not only for the density measurements but also for a consistent comparison of galaxy properties. We estimated the surface densities of red HAEs to be $0.067 \pm 0.013$ arcmin$^{-2}$ in BOSS1244 and $0.048 \pm 0.011$\,arcmin$^{-2}$ in BOSS1542, corresponding to red galaxy overdensities of $6.60 \pm 1.80$ and {\bf $4.45\pm1.18$}, respectively. These values are consistent with our previous results \citep{Zheng2021, Shi2021}.

For our selected red HAEs, we examine their morphologies in the F160W band (rest-frame $B$-band) utilizing HST observations. Figure~\ref{fig:fig4} shows the morphologies of red HAEs in BOSS1244 and BOSS1542. Among the 28 red HAEs in BOSS1244, 12 are covered by HST imaging, and the majority exhibit irregular morphologies or multiple components. In BOSS1542, 8 out of 19 red HAEs are detected by HST, also displaying irregular structures \citep{Liu2023, Golden-Marx2026}. 

Such disturbed and clumpy morphologies have also been reported for other populations of HAEs in protoclusters. For example, \cite{Naufal2023} found an enhanced abundance of disturbed and clumpy HAEs in protocluster environments, indicating that such morphologies may not be unique to red HAEs. Recent JWST observations have further revealed that clumpy and irregular structures are common among high-redshift galaxies, with prominent stellar substructures identified in both rest-frame UV and near-IR imaging \citep[e.g.,][]{Kalita2024,Kalita2025a,Kalita2025b,Sok2026,delaVega2026,Zhu2026}. The disturbed morphologies observed in our red HAEs may therefore be part of the broader population of clumpy and structurally disturbed galaxies at high redshift, rather than being a characteristic specific to the red HAE population.

\begin{figure*}[ht]
\setlength{\abovecaptionskip}{0pt}
\begin{center}
\includegraphics[trim=0mm 0mm 0mm 0mm,clip,height=0.33\textwidth]{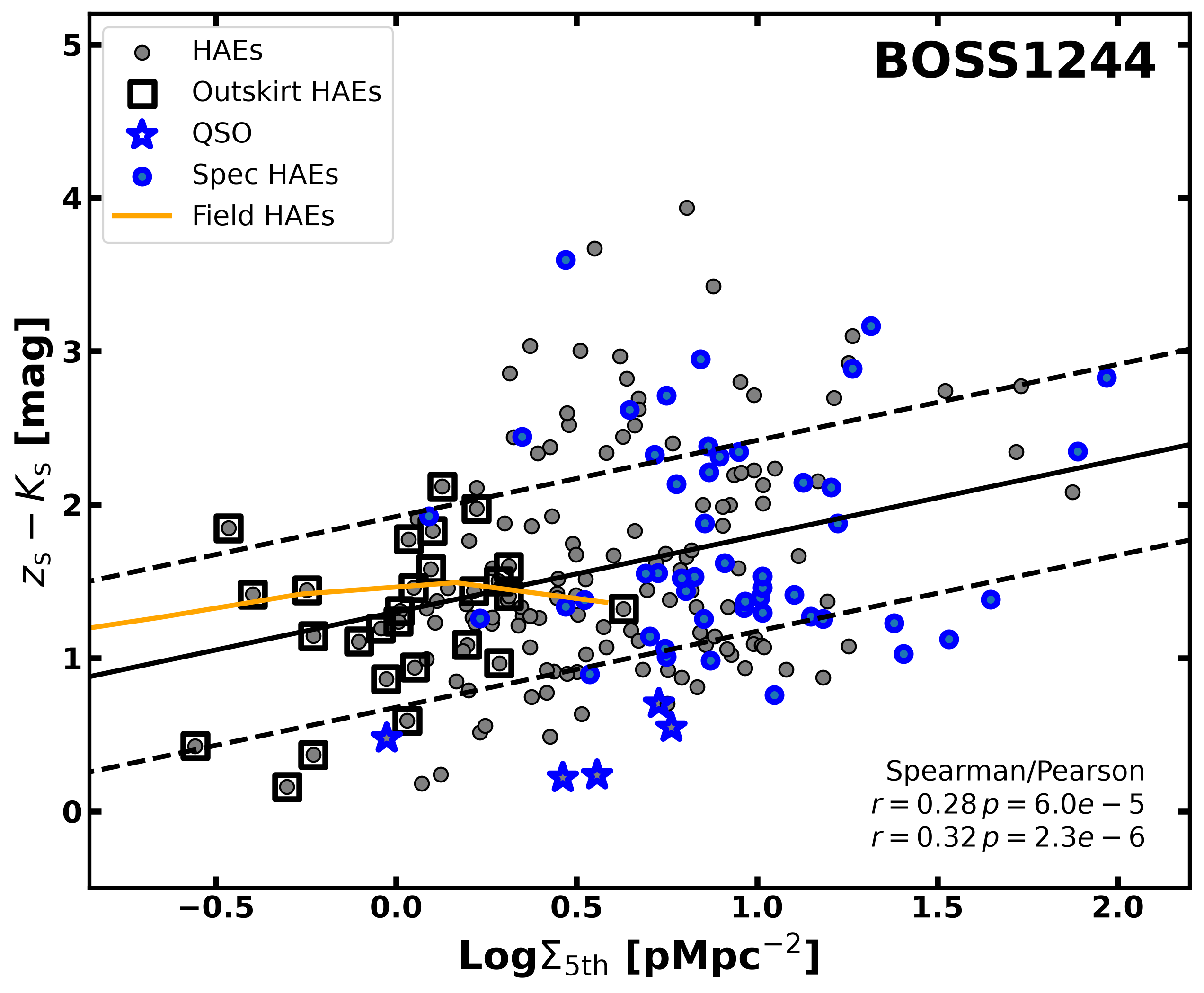}
\includegraphics[trim=0mm 0mm 0mm 0mm,clip,height=0.33\textwidth]{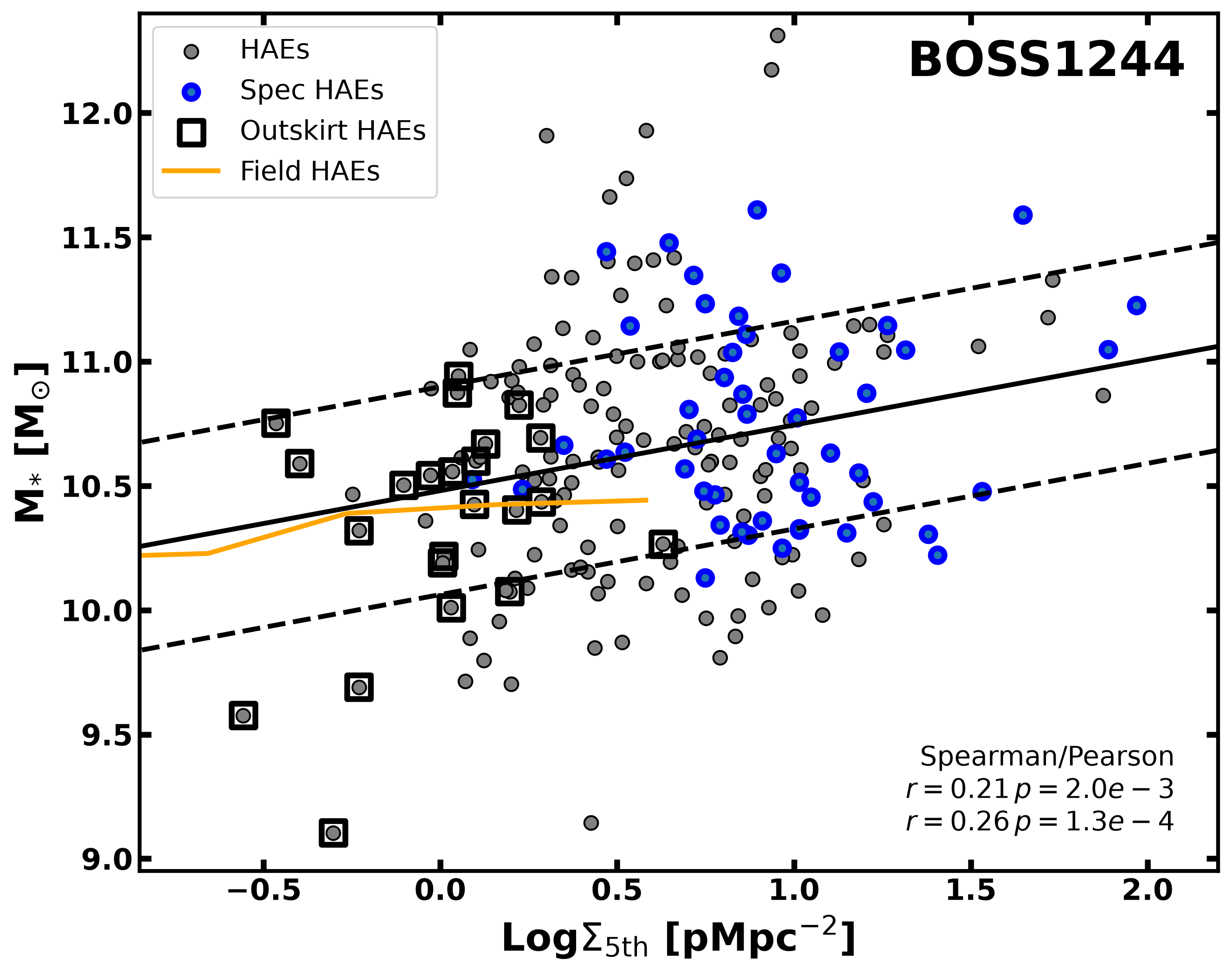}
\includegraphics[trim=0mm 0mm 0mm 0mm,clip,height=0.33\textwidth]{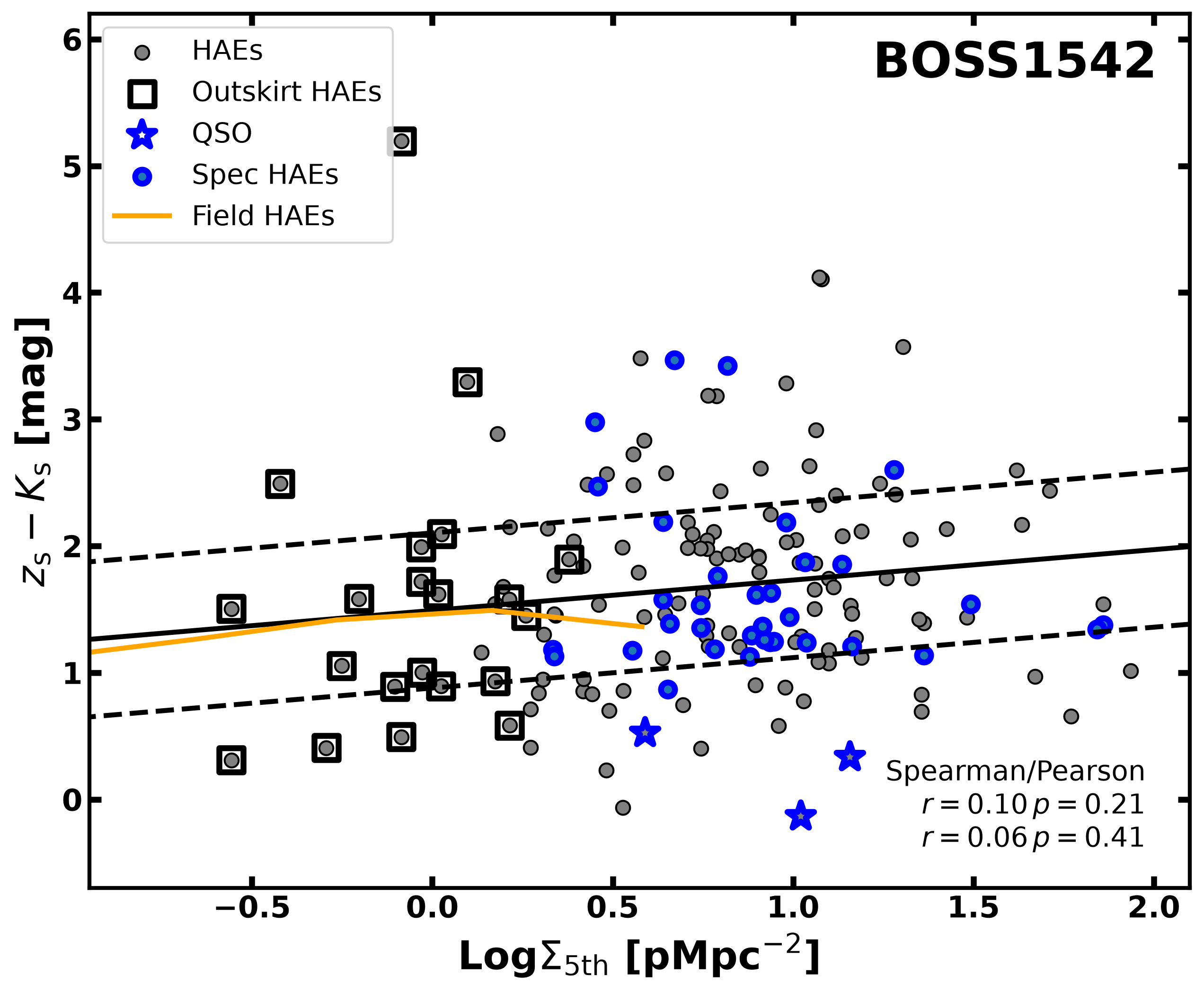}
\includegraphics[trim=0mm 0mm 0mm 0mm,clip,height=0.33\textwidth]{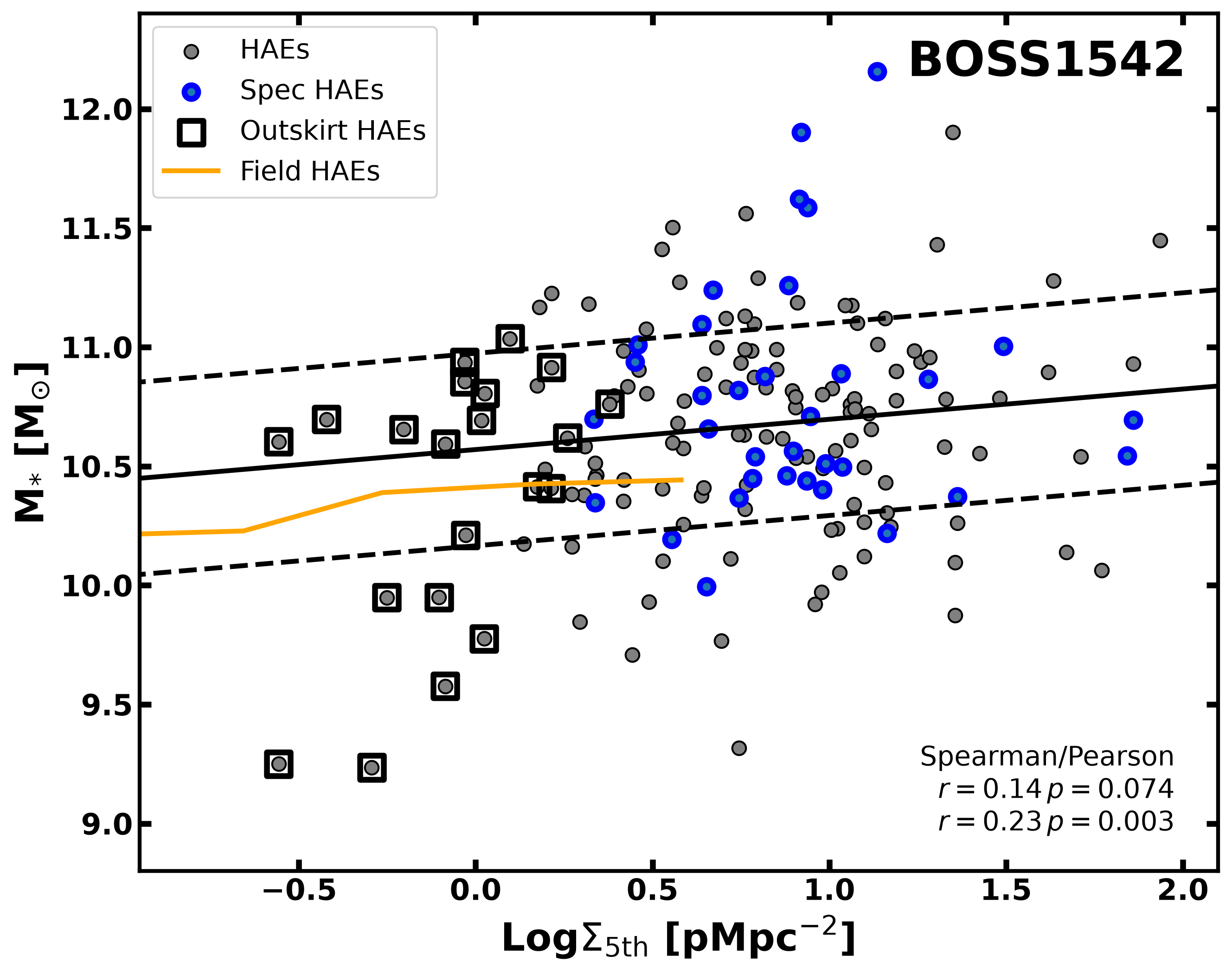}
\caption{Color (left) and stellar mass (right) as a function of the local density $\Sigma_{\rm 5th}$ for HAEs in BOSS1244 (upper) and BOSS1542 (bottom). Gray filled circles represent the NB-selected HAEs detected in both $z_{\rm s}$ and $K_{\rm s}$. The black squares are the outskirt HAEs, and the blue circles are spectroscopically confirmed HAEs. Blue stars show the BOSS quasars at $z=2.24$. Black solid line represents the best fit to the entire sample of  HAEs and black dashed lines show the 3$\sigma$ scatter. Orange solid lines give the relations between color/stellar mass and $\Sigma_{\rm 5th}$ for HAEs in general fields.} 

\label{fig:fig5}
\end{center}
\end{figure*}

Furthermore, we find that none of the red HAEs are located in the $Bz_{\rm s}K_{\rm s}$-selected passive galaxy region. This, together with their observed irregular or clumpy morphologies, is consistent with the interpretation that the majority of these red HAEs are still star-forming systems rather than quiescent galaxies. In BOSS1244-SW, an extremely massive red HAE is detected by JCMT 850\,$\mu m$, confirming that at least a subset of red HAEs are dust-obscured galaxies.  However, the HST coverage of our red HAE sample is incomplete, and the relatively limited spatial resolution and sensitivity of the available imaging prevent us from drawing firm conclusions about their detailed morphologies. Deeper and wider NIR imaging, particularly with JWST, will be valuable for systematically characterizing their stellar structures and determining whether the observed disturbed morphologies are associated with clump formation, galaxy interactions, or other physical processes.

\subsection{Local environments affect galaxy properties} 

\subsubsection{Color--density relation} 

The color--density relation has been extensively studied at $z\sim1$ \citep[e.g.,][]{Kodama2001,Cucciati2006,Cooper2010,Chuter2011,Quadri2012, Lemaux2019}. \cite{Koyama2014} identified HAEs and red-sequence galaxies in the overdensity 4C~65.22 at $z=1.5$, and reported a clear color--$\Sigma_{5\text{th}}$ relation. They found that galaxies in high-density regions are generally redder and dominated by passively evolving red galaxies. However, this color–density relation becomes much weaker and is almost absent when only HAEs are considered. It is worth noting that 4C~65.22 exhibits a substantially more prominent and well-established red sequence than those observed in BOSS1244 and BOSS1542. This difference provides important context when comparing the environmental dependence of galaxy colors among these systems, given their different evolutionary stages.

We investigate the relationship between the color $z_{\rm s}-K_{\rm s}$ and the local surface density $\Sigma_{5\text{th}}$ for HAEs in BOSS1244 and BOSS1542. We calculate both the Spearman rank and Pearson correlation coefficients, which characterize the strength of monotonic and linear associations, respectively. Following the empirical classification of \cite{Wuensch1996}, we describe the correlation with absolute coefficients of $0.00 \leq |r| < 0.20$ as very weak, $0.20 \leq |r| < 0.40$ as weak, $0.40 \leq |r| < 0.60$ as moderate, $0.60 \leq |r| < 0.80$ as strong, and $0.80 \leq |r| \leq 1.00$ as very strong. A correlation is considered statistically significant when $p \leq 0.05$, in which case the null hypothesis of no correlation is rejected at the 5\% significance level.

\begin{figure*}[ht]
\setlength{\abovecaptionskip}{0pt}
\begin{center}
\includegraphics[trim=0mm 0mm 0mm 0mm,clip,height=0.33\textwidth]{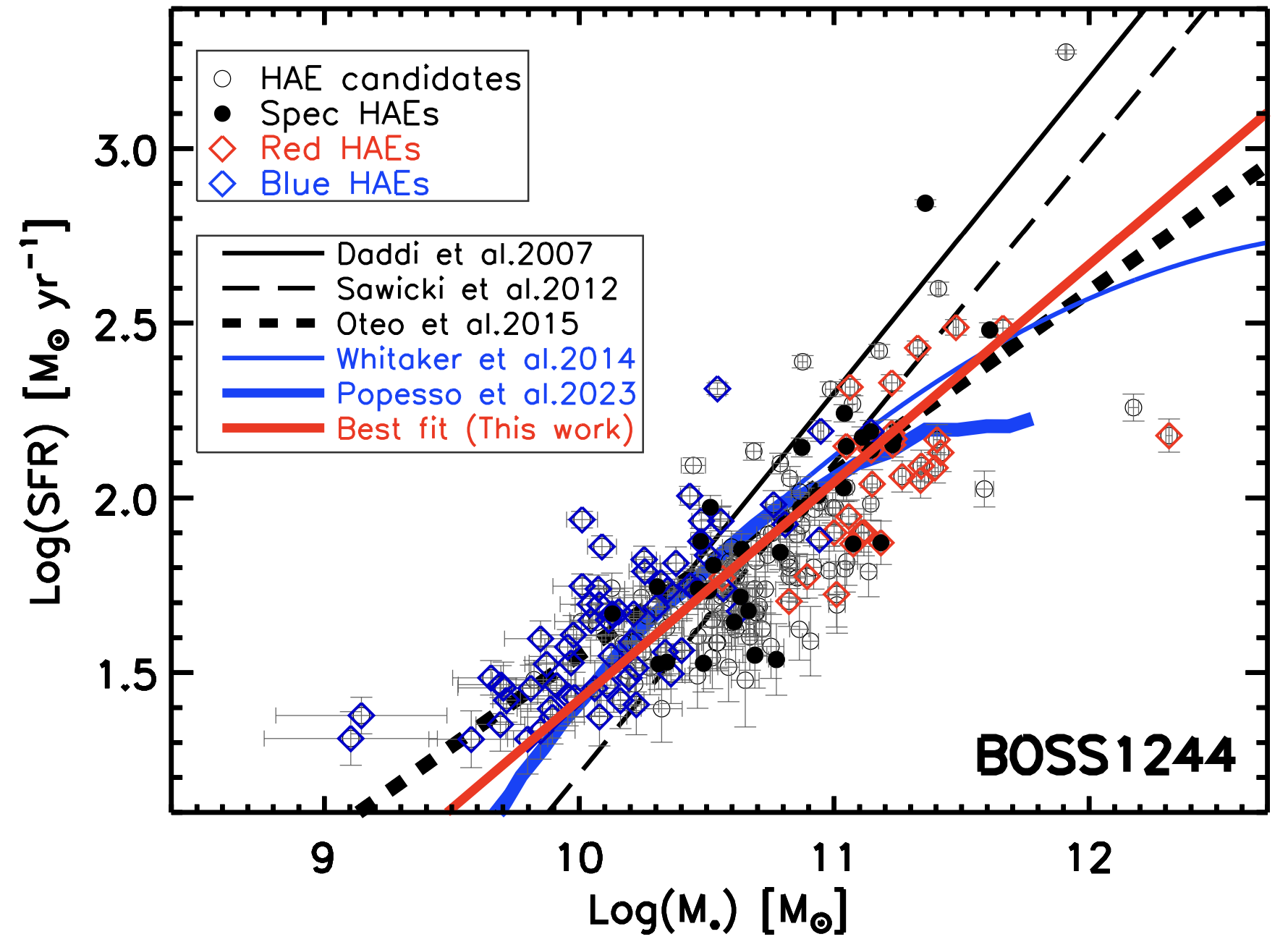}
\includegraphics[trim=0mm 0mm 0mm 0mm,clip,height=0.33\textwidth]{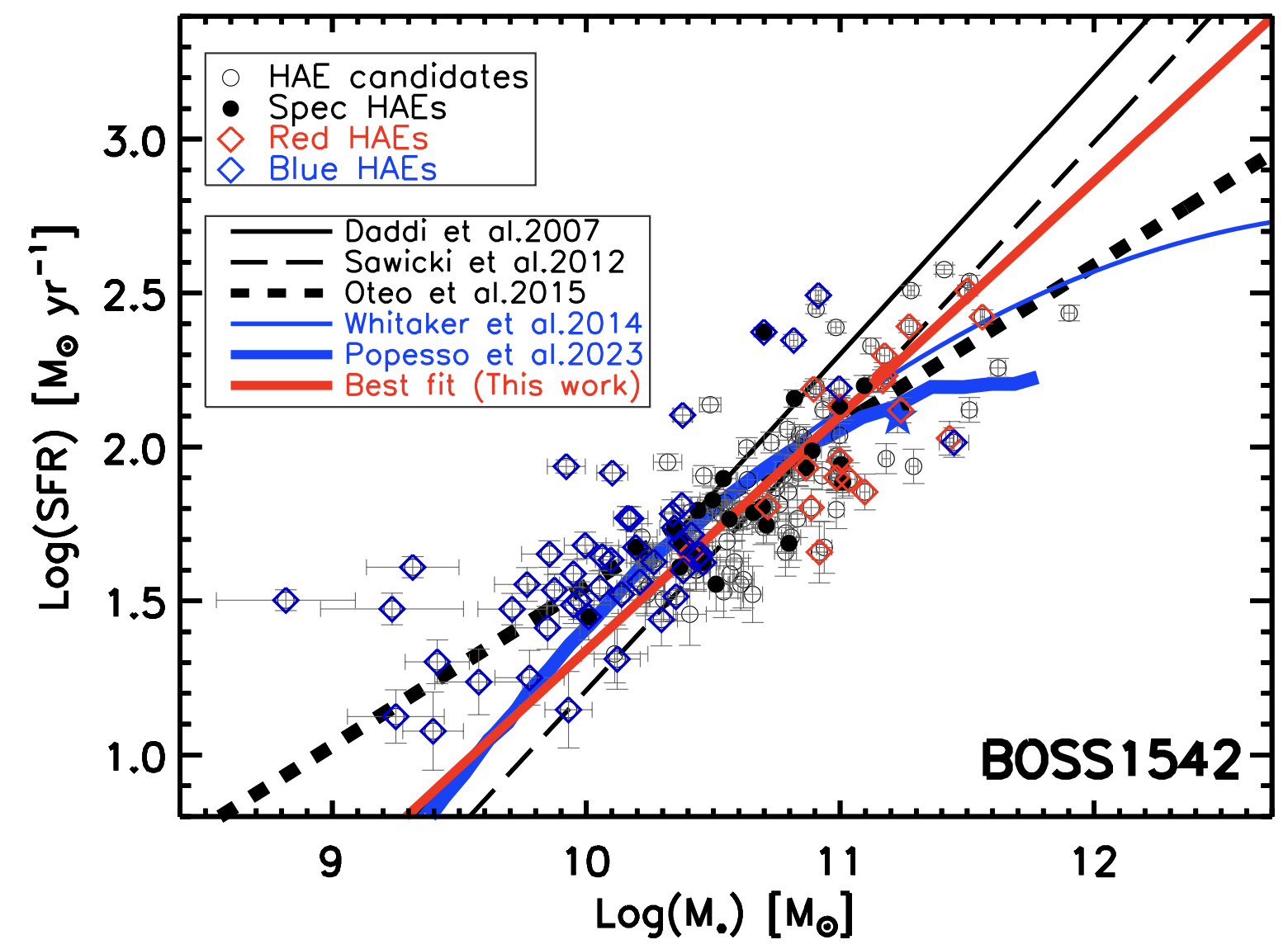}
\caption{The SFMS of HAEs in BOSS1244 (left) and BOSS1542 (right). Gray open circles are the HAE candidates, black filled circles show spectroscopically confirmed HAEs, red diamonds are red HAEs, and blue diamonds are blue HAEs. The black thin solid, black thin dashed, blue thin solid, black thick dotted, and blue thick solid lines represent SFMS relations at cosmic noon from the literature \citep{Daddi2007,Sawicki2012,Whitaker2014,Oteo2015,Popesso2023}. The red thick solid line represents the best-fit curve.} 
\label{fig:fig6}
\end{center}
\end{figure*}

For BOSS1244, we find Spearman and Pearson coefficients of 0.28 and 0.32, respectively, with corresponding $p$-values of $6.0\times10^{-5}$ and $2.3\times10^{-6}$. These results indicate a weak but statistically significant positive association between galaxy color and local density. In BOSS1542, the corresponding coefficients $r$ are 0.10  and 0.06, with corresponding $p$-values of 0.21 and 0.41, respectively, indicating very weak correlations that are not statistically significant.  For comparison, HAEs in the COSMOS field at $z=2.23$ show Spearman and Pearson coefficients of 0.14 and 0.11, with corresponding $p$-values of 0.004 and 0.02. The COSMOS results therefore indicate very weak but statistically significant associations. The difference in the correlation coefficients among the three fields should not, however, be interpreted as a direct measure of the relative strength of the environmental effect, because the coefficient depends on the distribution and dynamic range of the variables, as well as on sample size and intrinsic scatter. In particular, the similar local density ranges of BOSS1244 and BOSS1542 suggest that the stronger color--density association found in BOSS1244 cannot simply be attributed to a systematically denser environment. Instead, the difference may reflect the distinct evolutionary or dynamical states of the two protoclusters. Overall, these results suggest that a color--density association may already be present among HAEs at $z\sim2.2$, but its strength and statistical significance vary among individual environments. The left two panels of Figure~\ref{fig:fig5} illustrate these trends.

\subsubsection{Stellar mass--density relation} 

The possible dependence of the color--density relation on both stellar mass and local environment motivates us to further examine the relationship between stellar mass and $\Sigma_{5\text{th}}$. The right two panels of Figure~\ref{fig:fig5} show the relationship between stellar mass and $\Sigma_{5\text{th}}$ for HAEs in  BOSS1244, BOSS1542 and COSMOS.  
In BOSS1244, the Spearman and Pearson correlation coefficients are 0.21 and 0.26, with corresponding $p$-value of $2.0\times10^{-3}$ and $1.3\times10^{-4}$, respectively, indicating weak and statistically significant positive associations. In BOSS1542, the corresponding coefficients are 0.14 and 0.23, with corresponding $p$-values of 0.074 and 0.003. Thus, the Spearman test indicates a very weak association that does not reach our adopted significance threshold, whereas the Pearson test indicates a weak and statistically significant linear association. In COSMOS, the Spearman and Pearson coefficients are 0.22 and 0.15, with corresponding $p$-values of $5.7\times10^{-6}$ and 0.002, indicating weak and very weak, but statistically significant associations, respectively. 

Taken together, the three fields show evidence for modest associations between stellar mass and local density, although the strength and statistical significance depend on both the environment and the statistical measure adopted. As for the color--density relation, the different correlation coefficients should not be interpreted as a direct ranking of environmental effects among the fields. In particular, the presence of a statistically significant correlation in COSMOS does not imply that its environmental dependence is stronger or weaker than that in BOSS1244 solely on the basis of the correlation coefficient. Within BOSS1244, the simultaneous tendency for HAEs in denser regions to be redder and more massive is consistent with an association among local environment, galaxy color, and stellar mass. The contrast between BOSS1244 and BOSS1542, despite their broadly comparable density regimes, may again be related to differences in their evolutionary or dynamical states. These correlation analyses identify statistical associations but do not by themselves establish a causal relationship between environment, color, and stellar mass.

\begin{figure*}[ht]
\setlength{\abovecaptionskip}{0pt}
\begin{center}
\includegraphics[trim=0mm 0mm 0mm 0mm,clip,height=0.5\textwidth]{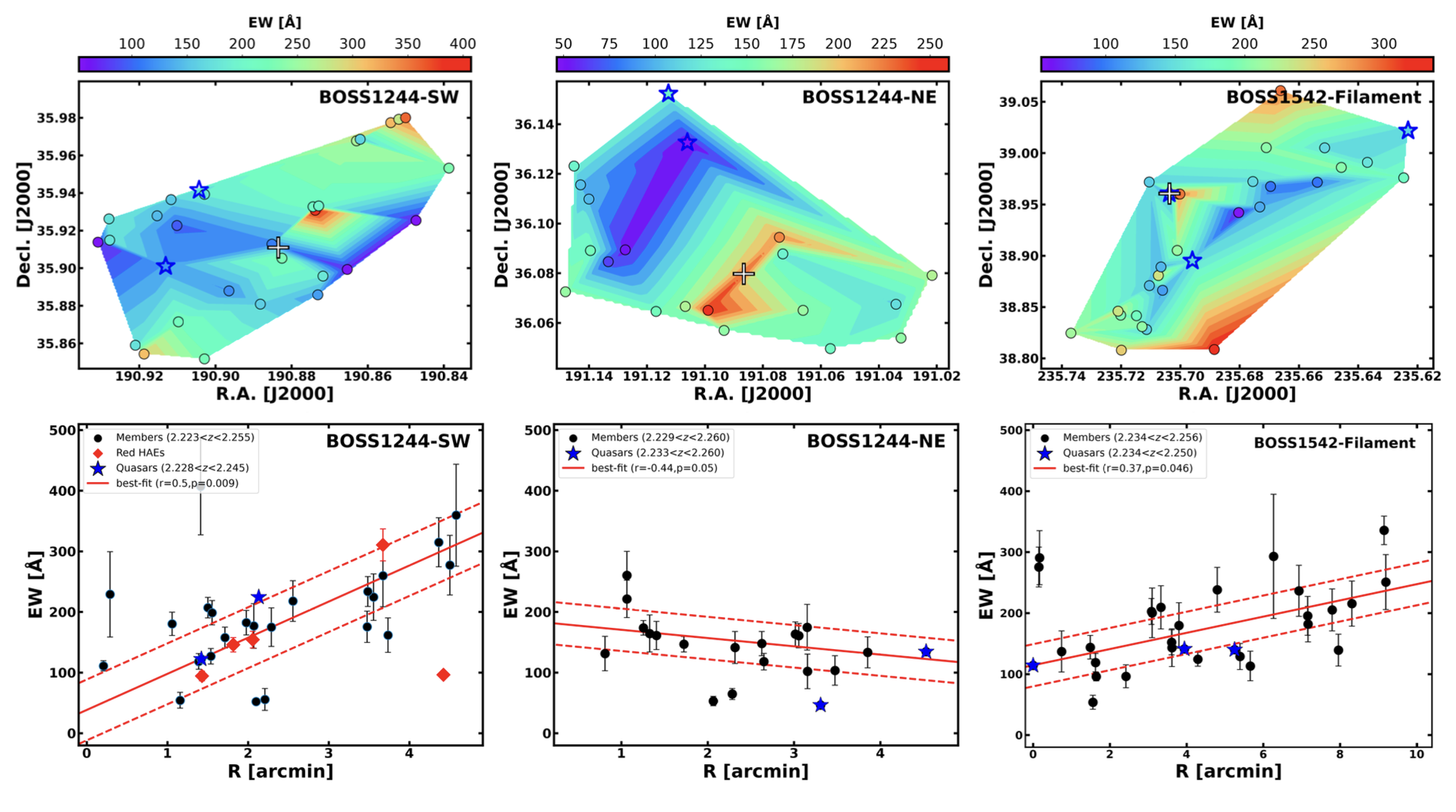}
\caption{ The first row, from left to right, shows the two-dimensional distribution maps of the H$\alpha$  EWs for spectroscopically confirmed
member HAEs in the BOSS1244-SW, BOSS1244-NE, and BOSS1542-Filament regions, respectively, with colors representing EW values. White cross in each panel marks the location of the central region. The second row, from left to right, represents the radial distributions of EWs for HAEs in the three regions, respectively. Red diamonds are the spectroscopically confirmed red HAEs. Black filled circles are HAEs, and blue stars are quasars. The red solid line indicates the best-fit curve, and the red dashed lines indicate the $1\sigma$ dispersion.} 
\label{fig:fig7}
\end{center}
\end{figure*}

\subsection{The star-forming main sequence in protocluster environments} \label{sec:spatial}

Numerous studies have established a correlation between SFR and stellar mass for distant star-forming galaxies \citep[e.g.,][]{Daddi2007, Elbaz2007, Santini2009, Sawicki2012, Whitaker2014, Oteo2015, Ciesla2017, Curtis-Lake2021, Popesso2023}, commonly referred to as the star-forming main sequence (SFMS). In Figure~\ref{fig:fig6}, we present the SFR--$M_{*}$ relation for HAEs in BOSS1244 and  BOSS1542. We further classified the HAEs in both fields into red ($z_{\rm s}-K_{\rm s}>2.5$) and blue ($z_{\rm s}-K_{\rm s}<1.2$) populations. The redder HAEs are found to predominantly occupy the massive end of the SFMS, exhibiting both higher SFRs and higher stellar masses than their blue counterparts. In the BOSS1244 field, the average SFRs for red and blue HAEs are 123 and $41\,M_{\odot}$ yr$^{-1}$, with corresponding average stellar masses of $1.68\times10^{11}\,M_{\odot}$ and $1.33\times10^{10}\,M_{\odot}$, respectively. In the BOSS1542 field, the average SFRs are 132 and $42\,M_{\odot}$\,yr$^{-1}$, and the average stellar masses are $1.25\times10^{11}\,M_{\odot}$ and $1.32\times10^{10}\,M_{\odot}$, respectively.

To achieve a completeness of $>80\%$  in detection, we restrict our fitting to HAEs with stellar masses exceeding $10^{10}\,M_{\odot}$. {In Figure~\ref{fig:fig6},} we find that the slope of the SFMS in BOSS1542 (slope: $0.76\pm0.01$) is steeper than that in BOSS1244 (slope: $0.62\pm0.01$). A steeper slope implies that SFR increases more rapidly with stellar mass. In other words, massive HAEs with $10^{10}\,M_{\odot}$ in BOSS1542 appear to be more actively forming stars than their counterparts in BOSS1244. This result is consistent with the estimate by \cite{Liu2023} that the merger rate of HAEs in BOSS1542 is slightly higher than that in BOSS1244, which may more effectively trigger gas collapse and star formation in massive galaxies. 

The different SFMS slopes may also be related to the distinct evolutionary and dynamical states of the two protoclusters. As discussed above, BOSS1542 exhibits a prominent filamentary structure, a relatively low velocity dispersion with 247\,km\,s$^{-1}$, and no statistically significant color--density or stellar mass--density relation. Together with its elevated merger activity and enhanced star formation, these properties are consistent with BOSS1542 being at an earlier stage of protocluster assembly, when galaxy growth and star formation are still actively proceeding. In contrast, BOSS1244 appears to be dynamically more evolved, with stronger environmental associations of galaxy color and stellar mass and a relatively shallower SFMS slope. The latter may indicate that the SFRs of some massive galaxies are becoming less responsive to increasing stellar mass, potentially reflecting the onset of reduced star-formation activity as the system evolves. This interpretation is consistent with \cite{Shi2024}, who found evidence that galaxies in BOSS1244 were undergoing inside-out quenching. The observed difference in SFMS slopes may therefore reflect, at least in part, the different stages of galaxy assembly and star-formation evolution in BOSS1542 and BOSS1244. However, the present results do not uniquely identify the physical mechanisms responsible, and processes such as changes in gas supply or AGN feedback may also contribute.

The HAEs in both fields broadly follow the SFMS reported in the literature, in the sense that galaxies with higher stellar masses exhibit higher SFRs. This suggests that, even in these high-redshift and high-density environments, the fundamental physics of star formation in galaxies remains consistent with that seen in typical field galaxies.  

\subsection{Spatial distributions of specific SFR traced by H$\alpha$ EW in overdense environments} \label{sec:ew}

EW is often used as a proxy of specific SFR, measuring a galaxy's SFR relative to its stellar mass \citep[e.g.,][]{Khostovan2024}. \cite{Shi2024} discovered a pair of massive quiescent galaxies in the densest region of BOSS1244-SW. Centered on this pair, the H$\alpha$ EWs of member galaxies in the BOSS1244-SW region gradually increase from the center to the outskirts, i.e., smaller EWs in the center and larger EWs in the outskirts. Motivated by this correlation, we carry out an analysis of  the radial distribution of H$\alpha$ EWs for HAEs in the BOSS1244-NE and BOSS1542-Filament regions. 

Figure~\ref{fig:fig7} presents the two-dimensional (2D) spatial distributions and one-dimensional (1D) radial profiles of the H$\alpha$ EWs for spectroscopically confirmed HAEs in the BOSS1244-SW, BOSS1244-NE, and BOSS1542-Filament regions. In BOSS1244-SW, the EW increases gradually from the central, high-density region centered on the brightest cluster member QG1 (R.A. = 190.8835000, Decl. = 35.9110222) toward the outskirts, consistent with the result reported in \cite{Shi2024}. In BOSS1244-NE, the enhanced H$\alpha$ EWs are concentrated in a relatively compact region, which we adopt as the reference center (R.A. = 191.08659, Decl. = 36.07973) for the radial analysis. In BOSS1542-Filament, the region with lower H$\alpha$ EWs is located close to the quasar. We therefore adopt the quasar (R.A. = 235.70368, Decl. = 38.960585) at $z= 2.242$ as the reference center and examine how the EW varies with increasing distance from it.

To quantify the spatial trends suggested by the 2D maps, we construct the corresponding 1D radial profiles. These profiles reveal positive H$\alpha$ EW gradients in BOSS1244-SW and BOSS1542-Filament, whereas the BOSS1244-NE profile is considerably flatter and shows only a weak, marginally significant negative trend. The Spearman correlation coefficients are $r=0.5$ ($p=0.009$) for BOSS1244-SW, $r=-0.44$ ($p=0.05$) for BOSS1244-NE, and $r=0.37$ ($p=0.046$) for BOSS1542-Filament. Thus, the H$\alpha$ EW increases with radius in BOSS1244-SW and BOSS1542-Filament, whereas it decreases with radius in BOSS1244-NE. These results confirm that the spatial EW gradients identified in the 2D maps are also reflected in the 1D radial distributions, with the EW enhancements or suppressions being relatively concentrated around the adopted central regions.

In the BOSS1244-SW region (see the left panels of  Figure~\ref{fig:fig7}), the radial distribution is derived from HAEs at $z=2.223-2.255$, and the resulting trend is fully consistent with that reported by \cite{Shi2024}. The positive EW gradient indicates that HAEs in the dense central region have lower H$\alpha$ EWs than those at larger radii. As discussed in \cite{Shi2024}, this trend may be related to enhanced interactions, gas depletion, or other processes that suppress star formation in the central region, while galaxies in the outskirts may retain more cold gas and sustain more active star formation. However, the present data do not allow us to uniquely distinguish among these possible mechanisms.

In the BOSS1244-NE region (see the middle panels of  Figure~\ref{fig:fig7}), the H$\alpha$ EW decreases with increasing radius from the adopted high-EW center, indicating that galaxies with enhanced EWs are spatially concentrated around this region. The negative correlation is relatively weak and is marginally significant ($r=-0.44$, $p=0.05$). This pattern may be associated with locally enhanced star-forming activity or a locally elevated gas supply. Nevertheless, it is difficult to determine whether the observed gradient is primarily driven by the galaxy environment. Projection effects and localized starburst activity may also contribute to the observed spatial pattern.

In the BOSS1542-Filament region (see the right panels of  Figure~\ref{fig:fig7}), the H$\alpha$ EW increases with increasing distance from the quasar-associated, low-EW region. This positive correlation ($r=0.37$, $p=0.046$) indicates that the lower-EW galaxies are preferentially concentrated around the quasar-associated region, while galaxies at larger distances tend to have higher EWs. This spatial pattern is consistent with the possibility that quasar-related processes may influence the star formation activity of galaxies. For example, quasar radiation could heat or ionize the surrounding gas and thereby suppress star formation, leading to lower H$\alpha$ EWs near the quasar. However, the current data do not provide sufficient evidence to establish a causal connection between the quasar and the observed EW gradient, and the contributions from the filamentary environment or projection effects cannot be ruled out. 

It is worth noting that in all three regions, HAEs located close to quasars tend to show smaller H$\alpha$ EWs. This common feature is consistent with the possibility that quasar activity may suppresses star formation in neighboring galaxies, possibly through UV/X-ray radiation heating or feedback mechanisms. However, this does not imply that the member galaxies evolve uniformly. The influence of quasar feedback is likely localized and may depend on the distance from the quasar and on the local density. Indeed, \cite{Golden-Marx2026} found that only the central AGN region near the quiescent BCGs in BOSS1244 shows strong evidence for an enhanced fraction of early-type galaxies, whereas the broader protocluster does not exhibit a uniform morphological transformation. Therefore, while quasar-related feedback may contribute to the observed EW gradients, the overall evolution of galaxies in these protoclusters appears spatially inhomogeneous. So far, our spectroscopic observations have been restricted to the high-density regions of HAEs in these two fields.  Future observations will cover more regions with spectroscopic data to identify protocluster members and examine the distribution of H$\alpha$ EWs.

\section{Discussions} \label{sec:discuss}

\subsection{Environmental trends in BOSS1244 and BOSS1542 } \label{sec:main-result}

Our analysis reveals a clear dichotomy in the environmental dependence of galaxy properties between the two protocluster fields, BOSS1244 and BOSS1542, at $z=2.24$. In BOSS1244, HAEs in higher-density regions exhibit redder $z_{\rm s}-K_{\rm s}$ color and systematically higher stellar masses than those in lower-density environments.  The trend is particularly strong in the BOSS1244-SW region, where a positive correlation between H$\alpha$ EWs of HAEs and radial distance is observed. In BOSS1244-NE, however, the H$\alpha$ EW decreases gradually from the center outward, with the center defined as the region of higher EW. This suggests that high-EW galaxies tend to cluster near a central area, possibly due to more recent star formation or a higher gas supply. However, with the current data, it is difficult to accurately determine whether environmental density dominates this gradient. The possibility of projection effects from large-scale structures or local starburst events cannot be ruled out. In the densest core of BOSS1244-SW, the presence of quiescent galaxies, the radial H$\alpha$ EW gradient observed across the entire member sample in \cite{Shi2024} (with lower EWs in the center and higher EWs in the outskirts), and red HAEs together 
support the interpretation that environmental quenching mechanisms (such as gas depletion, ram-pressure stripping, or AGN feedback) are already operating at this early epoch.
These results are consistent with the ``accelerated evolution'' scenario, wherein galaxies in overdense regions form earlier and more rapidly than their field counterparts.

In contrast, BOSS1542 shows weak or no color--density or mass--density relations. This field is dominated by an extended filamentary structure, and the H$\alpha$ EW increases with distance from a central quasar with relatively low EWs observed near the quasar. This radial trend suggests that radiative feedback or heating from the quasar may suppress star formation in its immediate vicinity, while star formation becomes more active in the outskirts. However, we cannot rule out projection effects or local starburst events as contributing factors. The absence of strong color--density and mass--density relations in BOSS1542 implies that this system is still in an early assembly phase, dominated by cold-gas accretion along the filament, with no widespread quenching yet established. This may also suggest that BOSS1542 is in an earlier stage of quenching compared to BOSS1244, where inside--out quenching has already been reported \citep{Shi2024}.

The SFMS for HAEs in both fields broadly follows the global SFR--$M_{\ast}$ relation, indicating that the fundamental scaling of star formation remains valid even in these extreme overdensities. Our SFMS results are broadly consistent with recent studies of other protoclusters at similar redshifts. \cite{Bertemes2026} reported slightly elevated star formation in the dense core of the protocluster ``The Step'' at $z\sim3$, although the inferred enhancement is subject to substantial uncertainties in the stellar masses. Similarly, \cite{Rinaldi2026} found that member galaxies in the $z\simeq1.84$ overdensity HUDF46 broadly follow the SFMS, with stellar and star-formation properties similar to those of coeval field galaxies. Together, these results suggest that star-forming galaxies in protocluster environments at $z\sim2-3$ generally remain close to the SFMS, although modest environmental differences in the normalization or slope may be present. Nevertheless, the SFMS slopes differ significantly between BOSS1244 and BOSS1542, with BOSS1542 exhibiting a steeper slope ($0.76\pm0.01$) than BOSS1244 ($0.62\pm0.01$). The steeper SFMS slope in BOSS1542 indicates that its massive HAEs are forming stars more vigorously than those in BOSS1244. This suggests that BOSS1542 may be in a younger, more gas-rich phase. In contrast, the shallower slope in BOSS1244, combined with the presence of red HAEs and quiescent galaxies, implies that some massive galaxies have already begun to quench or show reduced star formation efficiency due to gas depletion or feedback.

Taken together, these results demonstrate that environmental effects on galaxy properties are already in place at $z=2.24$, but their manifestation depends strongly on the evolutionary stage of the large-scale structure. BOSS1244 appears to be in a more mature phase, where quenching is starting to operate from the inside out, while BOSS1542 remains in a younger, gas-rich, assembling phase.

\subsection{Comparison with other protocluster studies } \label{sec:comparison}

Our findings extend previous work on environmental dependencies in protocluster galaxies at $z>2$ to higher overdensity regimes. The color–density relation detected in BOSS1244 is qualitatively similar to that reported by \citet{Koyama2013} for the PKS1138 (Spiderweb) protocluster at $z=2.16$, where HAEs in dense regions are redder in $J-K$ and have stellar masses exceeding $10^{11}\,M_{\odot}$. Likewise, studies of other protocluster candidates such as 4C+10.48 at $z=2.35$, 2QZ, J0200+015 and SSA13 at $z=2.23$ have found that HAEs in overdense environments tend to be more massive and more evolved than those in underdense regions \citep[e.g.,][]{Hatch2011,Matsuda2011}. Although these systems may evolve into galaxy clusters with total masses in the range of $10^{14}-10^{15}\,M_{\odot}$ \citep[e.g.,][]{Hatch2011,Shimakawa2018a}, this is lower than the total enclosed mass of $10^{16}\,M_{\odot}$ infereed for BOSS1244 and BOSS1542 \citep{Zheng2021,Shi2021}. Our BOSS1244 field, with a galaxy overdensity of $5.6\pm0.3$ over $\sim39$\,cMpc scales \citep{Zheng2021}, is substantially more massive and more overdense than most previously studied systems, reinforcing the trend that environmental effects become stronger in the most massive protoclusters.

The contrast between BOSS1244 and BOSS1542 echoes the well-known dichotomy between PKS1138 at $z=2.16$ and USS1558 at $z=2.53$ \citep{Shimakawa2018a, Shimakawa2018b}. \cite{Shimakawa2018a} found that HAEs in USS1558 exhibit enhanced star formation in dense group cores, whereas \citet{Shimakawa2018b} reported no strong environmental dependence in PKS1138. They interpreted this as reflecting different evolutionary stages, USS1558 is a younger, more fragmented protocluster system, while PKS1138 is more mature with an established red sequence.  BOSS1542 with the filamentary morphology and low velocity dispersion \citep[$247\pm32\ {\rm km\,s^{-1}}$][]{Shi2021}, resembles USS1558, while BOSS1244 with multiple kinematic components and quiescent galaxies resembles PKS1138. This parallel supports the idea that the environmental dependence of galaxy properties is intimately linked to the assembly history and dynamical state of the protocluster.

The different SFMS slopes between BOSS1244 and BOSS1542 also find counterparts in previous work. The steeper slope in BOSS1542 indicates that massive HAEs are more actively forming stars, similar to the behavior seen in USS1558, where the densest subgroups show enhanced SFR across the main sequence. The shallower slope in BOSS1244 is consistent with PKS1138, where central HAEs are massive but do not show significant SFR enhancement at fixed mass, likely due to the onset of quenching. \citet{Liu2023} reported a slightly higher merger rate in BOSS1542, which could trigger more efficient gas collapse and star formation, further explaining its steeper SFMS and lack of strong color--density relation. In addition, 
\cite{Golden-Marx2026} found a stronger, albeit still weak, morphology--density relation in BOSS1244 compared to BOSS1542, with a larger fraction of early-type galaxies in its denser regions. This supports the interpretation that BOSS1244 is in a more advanced evolutionary stage, where morphological transformation and quenching are already underway, whereas BOSS1542 remains in an earlier, gas-rich phase.

\subsection{Comparison with theoretical simulations} \label{sec:simulations}

Our observational results can be directly compared with recent predictions from cosmological simulations. \cite{Andrews2025} examined protocluster galaxy properties using the IllustrisTNG simulations, specifically investigating Ly$\alpha$ emitters (LAEs) as tracers of large-scale structure. They found that protocluster galaxies follow the same SFR--$M_{*}$ scaling relation as average field galaxies, but a larger fraction have undergone major mergers and show star formation enhanced by $\sim60$\%. Our finding that HAEs in both BOSS1244 and BOSS1542 broadly follow the SFMS, but with different slopes between the two systems, is qualitatively consistent with these simulation predictions.  \cite{Andrews2025} also found that protocluster galaxies begin to quench much earlier ($z=0.8-1.6$) than field galaxies ($z=0.5-0.9$). The presence of quiescent galaxies in BOSS1244-SW at $z = 2.24$ supports this prediction, indicating that quenching is already underway in the densest protocluster cores at this early epoch \citep{Shi2024}.

\cite{Esposito2025} analyzed 14 DIANOGA protoclusters at $z = 2.2$. The total masses (${M}_{200, z=0}$) of these simulated protoclusters range from $1.5\times10^{14}$ to $3.2\times10^{15}\,M_{\odot}$, with a median of $1.6\times10^{15}\,M_{\odot}$. This range is comparable to the total enclosed masses inferred for the individual components BOSS1244-SW, BOSS1244-NE and BOSS1542-Filament \citep[$\sim10^{15}\,M_{\odot}$ each;][]{Shi2021}, although the entire BOSS1244 and BOSS1542 systems are more massive ($\sim10^{16}\,M_{\odot}$ over $\sim39$ cMpc). They found that the simulated protoclusters show an excess of high-mass galaxies with $M_{*}>10^{10}\, M_{\odot}$ in massive halos compared to the cosmological box, indicating accelerated evolution before infall. Our detection of a mass--density relation in BOSS1244, where HAEs in higher-density regions are significantly more massive, is consistent with this prediction. \cite{Esposito2025} predicted a weak suppression of SFRs in protocluster galaxies ($\sim$0.05\,dex) compared to the cosmological box, which increases to $\sim$0.25\,dex in massive halos, reflecting suppressed cold gas reservoirs. This is qualitatively consistent with the less active star formation we observe in BOSS1244 compared to BOSS1542, particularly in the densest regions.  The cold gas depletion scenario proposed by \cite{Esposito2025} to explain the SFR suppression in massive halos may also apply to BOSS1244. The presence of quiescent galaxies in the SW region and the redder colors of HAEs in higher-density regions suggest that gas depletion and quenching mechanisms such as AGN feedback, ram-pressure stripping, or starvation, are becoming effective in the densest protocluster cores at $z = 2.24$.

Furthermore, \cite{Shimakawa2018a} proposed a three-stage scenario  with growing, maturing and declining phase for cluster formation based on observations of USS1558 at $z = 2.53$ and PKS1138 at $z = 2.16$. Our results for BOSS1244 and BOSS1542 at $z = 2.24$ provide strong support for this scenario. BOSS1542 exhibits characteristics of the growing phase (filamentary structure, low velocity dispersion, weak environmental trends, steep SFMS), while BOSS1244 exhibits those of the maturing phase (strong color–density relation, red HAEs in dense cores, quiescent galaxies, and radial EW gradients indicative of inside-out quenching). This demonstrates that protoclusters at the same cosmic epoch can be in dramatically different evolutionary stages, depending on their assembly history and initial overdensities.

\section{Summary} \label{sec:summary}

In this work, we have analyzed the physical properties of star-forming HAEs in two extremely overdense structures at $z=2.24$, BOSS1244 and BOSS1542, using existing multi-wavelength imaging data from optical to near-infrared bands, with the aim of understanding the dependence of galaxy properties on environment. The main results are summarized as follows:

\begin{enumerate}

\item We examined the purity of our NB-selected HAE sample using the color-color distributions to exclude low- or high-redshift contaminants. About $84\%$ (151/180) of HAEs in BOSS1542 satisfy both $Bz_{\rm s}K_{\rm s}$ and $UBz_{\rm s}$ criteria. Accounting for color transformations, approximately $93\%$ (207/223) of HAEs in BOSS1244 also meet the $Uz_{\rm s}K_{\rm s}$ selection criteria. Nearly all spectroscopically confirmed HAEs satisfy the color-selection criteria, except for a few objects that lack detections in the required bands or have large photometric uncertainties, indicating that the color-selection criteria are highly consistent with the spectroscopic confirmation. This further confirms that most of the emission-line galaxies selected via the NB method are robust and lie within the same redshift range, reinforcing that both BOSS1244 and BOSS1542 represent genuinely dense environments. 

\item HAEs in denser regions of both protoclusters tend to be redder and more massive than those in low-density regions. This trend is more pronounced in BOSS1244, whereas BOSS1542 shows a flatter relation. This indicates that color--density and mass--density relations are already in place at $z=2.24$ and suggests that BOSS1244 and BOSS1542 are in different evolutionary stages. The presence of a subset of HAEs with red colors $(z_{\rm s}-K_{\rm s}>2.5)$ and high stellar masses $(>10^{11}\,M_{\odot})$ provides evidence that even in a vigorous star-forming phase, cluster galaxies had already formed a significant fraction of their stellar mass by $z=2-3$.

\item Red HAEs are preferentially distributed in the high-density regions of BOSS1244 and BOSS1542. They are brighter in the $K_{\rm s}$-band and have higher stellar masses and SFRs than blue HAEs. Compared to red HAEs in general fields at the same epoch, the overdensities of red HAEs in the BOSS1244 and BOSS1542 fields are 6.6 and 4.5, respectively, which are consistent with the HAE overdensity estimated for the entire fields. HST NIR imaging in F160W reveals that some red HAEs show irregular structures, suggesting they are still undergoing mass assembly. Furthermore, in the SW region of BOSS1244, an extremely massive red HAE $(>10^{12}\,M_{\odot})$ is detected by JCMT $850\,\mu$m, suggesting that at least a subset of red HAEs are dust-obscured galaxies.

\item The HAEs in BOSS1244 and BOSS1542 broadly follow the SFMS. However, the slope of SFMS in BOSS1542 (slope: $0.76\pm0.01$) is steeper than that in BOSS1244 (slope: $0.62\pm0.01$), suggesting that massive HAEs in BOSS1542 appear to be more actively forming stars than their counterparts in BOSS1244. This further suggests that the two systems are at different evolutionary stages.

\item The three overdense regions exhibit non-uniform radial H$\alpha$ EW gradients. The EW increases outward in BOSS1244-SW and BOSS1542-Filament, but decreases outward in BOSS1244-NE (though only weakly and marginally significantly). In addition, HAEs near quasars tend to show lower EWs across all three regions, which is consistent with a possible role for quasar-related feedback or radiation heating in suppressing star formation in neighboring galaxies. However, the current data cannot establish a causal connection, and contributions from the filamentary environment, projection effects, and localized starburst activity cannot be excluded. The independent role of local environmental density therefore remains uncertain and will require more extensive spectroscopic observations.

\end{enumerate}
 
 Our findings have broader implications for understanding how environment drives galaxy evolution at high redshift. The strong environmental dependencies in BOSS1244 confirm that the ``accelerated evolution'' paradigm where galaxies in overdense regions form earlier and more rapidly, is already operating at $z=2.24$. The presence of massive red HAEs and quiescent galaxies in the densest cores suggests that a significant fraction of the stellar mass in present-day cluster galaxies was already in place by $z\sim2-3$, consistent with earlier studies of PKS1138 \citep{Kodama2007}. The radial EW gradient in BOSS1244-SW provides direct observational evidence for inside-out quenching, supporting theoretical models of cluster formation.

However, several caveats should be considered when interpreting our results. First, our HAE sample may be incomplete at the faint end due to the flux limit of the NB observations, potentially biasing our sample toward galaxies with high H$\alpha$ EWs. Second, the stellar mass estimates rely on single-color ($K_{\rm s}$-band) measurements, which may introduce systematic uncertainties compared to full SED fitting. Nevertheless, \cite{Koyama2013a} demonstrated that this method yields consistent results with SED fitting for HAE samples at these redshifts. Third, the dust extinction correction applied to SFR estimates uses an empirical mass-dependent relation calibrated for local galaxies, which may not perfectly apply to high-redshift protocluster environments. However, this approach has been validated at $z\sim1.5-2$ in previous studies \citep{Hayashi2011,Sawicki2012,Koyama2013,Koyama2014}.  Fourth, spectroscopic confirmation of HAEs is currently incomplete, and future observations with deep multi-wavelength spectroscopy will be essential to securely determine the redshifts and derive accurate physical parameters for individual galaxies. 

Looking ahead, larger statistical samples of protoclusters at similar redshifts are urgently needed to disentangle the roles of assembly history, cold-gas accretion, AGN feedback, and quenching. The stark contrast between BOSS1244 and BOSS1542 highlights that protoclusters at the same cosmic epoch can be in vastly different evolutionary phases, and that environmental effects on galaxy evolution are not monolithic but depend critically on the specific large-scale structure context. Future work combining deep multi-wavelength imaging, spectroscopy, and high-resolution simulations will be crucial to fully unravel the complex interplay between environment and galaxy evolution in the early Universe.

\begin{acknowledgements}
    
We thank the anonymous referee for the valuable suggestions that significantly improved the manuscript. This work is supported by the start-up funding of the Anhui University of Science and Technology (2024jrc104), the National Science Foundation of China (grant no. 12303015 and 12233005), the National Science Foundation of Jiangsu Province (BK20231106) and the China Manned Space Program with grant No. CMS-CSST-2025-A20. XZZ acknowledges support by the National Key Research and Development Program of China (2023YFA1608100),  and in part by Office of Science and Technology, Shanghai Municipal Government (grant Nos. 24DX1400100, ZJ2023-ZD-001). HD acknowledges support from the Agencia Estatal de Investigación del Ministerio de Ciencia, Innovación y Universidades (MCIU/AEI) under grant (Construcción de cúmulos de galaxias en formación a través de la formación estelar oscurecida por el polvo) and the European Regional Development Fund (ERDF) with reference (PID2022-143243NB-I00/DOI:10.13039/501100011033).

\end{acknowledgements}

\bibliography{references}

\begin{appendix}
\onecolumn
\section{Red HAEs in BOSS1244 and BOSS1542}

Tables~\ref{tab:tab3} and \ref{tab:tab4} list the physical properties of red HAEs in BOSS1244 and BOSS1542, respectively.

\begin{longtable}[c]{cccccc}
\caption{The properties of 28 red HAEs with $z_{\rm s}-K_{\rm s}>2.5$ in BOSS1244.} \\
\label{tab:tab3} \\
\hline
\hline
R.A. & Decl. & $z_{\rm s}-K_{\rm s}$ & $\log(M_{*})$ & SFR & Redshift \\
 & & [mag] & [$M_{\sun}$] & [$M_{\sun}$\,yr$^{-1}$] & spec-$z$ \\
\hline
190.93154 & 35.838423 & $2.82\pm0.11$ & $11.23\pm0.02$ & $152.93\pm13.58$ & $\dots$ \\
190.99625 & 35.844489 & $2.69\pm0.16$ & $11.01\pm0.03$ & $52.97\pm11.45$ & $\dots$ \\
190.85396 & 35.846494 & $2.97\pm0.19$ & $11.00\pm0.03$ & $79.56\pm9.71$ & $\dots$ \\
190.93837 & 35.849224 & $2.62\pm0.05$ & $11.48\pm0.01$ & $307.41\pm15.74$ & $2.223$ \\
190.92214 & 35.851494 & $2.74\pm0.13$ & $11.06\pm0.02$ & $207.21\pm10.47$ & $\dots$ \\
190.91718 & 35.852642 & $2.77\pm0.10$ & $11.33\pm0.02$ & $268.28\pm12.38$ & $\dots$ \\
190.91874 & 35.854295 & $2.83\pm0.16$ & $11.23\pm0.02$ & $213.21\pm11.48$ & $2.246$ \\
190.88740 & 35.863770 & $2.62\pm0.16$ & $11.06\pm0.03$ & $88.41\pm9.60$ & $\dots$ \\
190.91472 & 35.867108 & $2.66\pm0.18$ & $10.89\pm0.03$ & $59.59\pm8.52$ & $\dots$ \\
191.00367 & 35.872738 & $3.04\pm0.13$ & $11.34\pm0.02$ & $112.27\pm13.29$ & $\dots$ \\
191.05200 & 35.886630 & $2.66\pm0.21$ & $10.82\pm0.03$ & $50.57\pm8.37$ & $\dots$ \\
190.96418 & 35.910781 & $2.80\pm0.14$ & $12.31\pm0.01$ & $150.63\pm16.50$ & $\dots$ \\
190.96409 & 35.912358 & $3.42\pm0.30$ & $11.09\pm0.03$ & $77.05\pm11.00$ & $\dots$ \\
190.91350 & 35.922077 & $2.70\pm0.12$ & $11.15\pm0.02$ & $109.52\pm10.33$ & $\dots$ \\
190.91011 & 35.922661 & $3.85\pm0.29$ & $11.23\pm0.02$ & $141.69\pm11.35$ & 2.225 \\
190.91401 & 35.926802 & $3.10\pm0.26$ & $11.11\pm0.03$ & $79.19\pm9.88$ & $\dots$ \\
190.91533 & 35.927837 & $3.17\pm0.21$ & $11.05\pm0.02$ & $140.35\pm9.86$ & 2.227 \\
190.91166 & 35.936436 & $2.89\pm0.16$ & $11.15\pm0.02$ & $136.90\pm10.36$ & 2.226 \\
190.84567 & 35.948171 & $2.61\pm0.33$ & $10.56\pm0.04$ & $58.90\pm6.68$ & $\dots$ \\
190.91737 & 35.965058 & $2.86\pm0.10$ & $11.34\pm0.02$ & $123.35\pm12.84$ & $\dots$ \\
190.95979 & 36.031495 & $2.52\pm0.04$ & $11.66\pm0.02$ & $305.66\pm18.35$ & $\dots$ \\
191.10081 & 36.043873 & $3.67\pm0.18$ & $11.40\pm0.01$ & $121.81\pm12.21$ & $\dots$ \\
191.11684 & 36.064582 & $2.71\pm0.11$ & $11.23\pm0.02$ & $147.21\pm10.98$ & 2.245 \\
190.97119 & 36.076117 & $2.60\pm0.07$ & $11.40\pm0.13$ & $146.61\pm14.63$ & $\dots$ \\
191.12745 & 36.089387 & $2.95\pm0.13$ & $11.18\pm0.02$ & $74.35\pm10.65$ & 2.251 \\
191.13083 & 36.135860 & $3.00\pm0.14$ & $11.27\pm0.01$ & $115.08\pm11.84$ & $\dots$ \\
190.88846 & 36.161365 & $2.71\pm0.11$ & $11.12\pm0.02$ & $79.68\pm12.38$ & $\dots$ \\
190.88224 & 36.163351 & $2.52\pm0.07$ & $11.42\pm0.01$ & $134.58\pm16.81$ & $\dots$ \\
\hline\hline
\end{longtable}

\begin{longtable}[c]{cccccc}
\caption{The properties of 19 red HAEs with $z_{\rm s}-K_{\rm s}>2.5$ in BOSS1542.} \\
\label{tab:tab4} \\
\hline
\hline
R.A. & Decl. & $z_{\rm s}-K_{\rm s}$ & $\log(M_{*})$ & SFR & Redshift \\
 & & [mag] & [$M_{\sun}$] & [$M_{\sun}$\,yr$^{-1}$] & spec-$z$ \\
\hline
235.71696 & 38.803202 & $3.18\pm0.13$ & $11.10\pm0.02$ & $71.32\pm9.58$ & $\dots$ \\
235.81871 & 38.805318 & $2.91\pm0.11$ & $11.18\pm0.02$ & $197.82\pm10.60$ & $\dots$ \\
235.76035 & 38.820478 & $3.81\pm0.51$ & $10.71\pm0.04$ & $63.97\pm7.41$ & $\dots$ \\
235.71590 & 38.826737 & $2.60\pm0.10$ & $10.90\pm0.02$ & $153.98\pm8.73$ & $\dots$ \\
235.72011 & 38.842011 & $2.60\pm0.15$ & $10.87\pm0.03$ & $85.46\pm8.34$ & 2.237 \\
235.71825 & 38.859269 & $3.30\pm0.18$ & $11.00\pm0.02$ & $90.44\pm9.14$ & $\dots$ \\
235.81998 & 38.868539 & $3.64\pm0.28$ & $10.99\pm0.03$ & $79.53\pm8.97$ & $\dots$ \\
235.86367 & 38.878782 & $2.88\pm0.10$ & $11.17\pm0.02$ & $163.99\pm13.35$ & $\dots$ \\
235.75413 & 38.894294 & $3.19\pm0.05$ & $11.56\pm0.01$ & $264.15\pm13.82$ & $\dots$ \\
235.86351 & 38.895363 & $2.74\pm0.35$ & $10.41\pm0.07$ & $44.97\pm7.22$ & $\dots$ \\
235.69422 & 38.898528 & $3.57\pm0.12$ & $11.43\pm0.01$ & $106.57\pm13.43$ & $\dots$ \\
235.50053 & 38.932825 & $3.30\pm0.20$ & $11.03\pm0.02$ & $78.33\pm9.72$ & $\dots$ \\
235.67551 & 38.972240 & $3.47\pm0.21$ & $11.24\pm0.02$ & $131.40\pm12.37$ & 2.237 \\
235.63688 & 38.990952 & $2.63\pm0.10$ & $11.00\pm0.02$ & $135.14\pm10.83$ & 2.242 \\
235.64352 & 39.010000 & $2.57\pm0.12$ & $10.89\pm0.04$ & $63.46\pm8.93$ & $\dots$ \\
235.79282 & 39.026432 & $3.48\pm0.15$ & $11.27\pm0.02$ & $245.82\pm11.17$ & $\dots$ \\
235.78591 & 39.073263 & $2.73\pm0.05$ & $11.50\pm0.01$ & $321.54\pm13.30$ & $\dots$ \\
235.73577 & 39.093992 & $2.63\pm0.06$ & $11.18\pm0.01$ & $170.16\pm11.32$ & $\dots$ \\
235.73469 & 39.100935 & $2.84\pm0.16$ & $10.92\pm0.03$ & $45.49\pm10.46$ & $\dots$ \\
\hline
\hline
\end{longtable}

 \end{appendix}

\end{document}